\documentclass[reprint, preprintnumbers, aps, prd, floatfix, superscriptaddress, nofootinbib]{revtex4-2}

\usepackage{amssymb, amsmath, physics, graphicx, xcolor, hyperref, mathtools, orcidlink}
\usepackage{fontawesome} % icon github links 

\newcommand\sYlm[3]{\ensuremath{{}_{#1}Y^{#2 #3}}}

\newcommand{\IoA}{Institute of Astronomy, University of Cambridge, Madingley Road, Cambridge, CB3 0HA, UK}
\newcommand{\KICC}{Kavli Institute for Cosmology, University of Cambridge, Madingley Road, Cambridge, CB3 0HA, UK}
\newcommand{\DAMTP}{Department of Applied Mathematics and Theoretical Physics, Centre for Mathematical Sciences, University of Cambridge, Wilberforce Road, Cambridge, CB3 0WA, UK}

\begin{document}

\title{Quasinormal modes from numerical relativity with Bayesian inference}

\author{Richard Dyer\,\orcidlink{0009-0008-3720-6092}}
\email{richard.dyer@ast.cam.ac.uk}
\affiliation{\IoA}

\author{Christopher J.\ Moore\,\orcidlink{0000-0002-2527-0213}}
\email{christopher.moore@ast.cam.ac.uk}
\affiliation{\IoA} \affiliation{\DAMTP} \affiliation{\KICC}

\date{\today}

\begin{abstract}
    Numerical relativity (NR) enables the study of physics in strong and dynamical gravitational fields and provides predictions for the gravitational-wave signals produced by merging black holes. 
    Despite the impressive accuracy of modern codes, the resulting waveforms inevitably contain numerical uncertainties. 
    Quantifying these uncertainties is important, especially for studies probing subdominant or nonlinear effects around the merger and ringdown.
    This paper describes a flexible Gaussian-process model for the numerical uncertainties in all the spherical-harmonic waveform modes across a state-of-the-art catalog of NR waveforms and a highly efficient procedure for sampling the posteriors of quasinormal mode models without the need for expensive Markov chain Monte Carlo. 
    The Gaussian-process model is used to define a likelihood function which allows many Bayesian data analysis techniques - already widely used in the analysis of experimental gravitational wave data - to be applied to NR waveforms as well. 
    The efficacy of this approach is demonstrated by applying it to the analysis of quasinormal modes in Cauchy-characteristic evolved waveforms. 
\end{abstract}

\maketitle

%%%
\section{Introduction} \label{sec:intro}

The final stage of a black hole (BH) merger, known as the ringdown, is associated with the remnant object settling into its final Kerr state. 
During this stage, perturbation theory predicts the existence of damped sinusoidal oscillations, known as quasinormal modes (QNMs), at specific frequencies as a prominent component of the gravitational wave (GW) signal. 
These QNM frequencies contain information about the remnant BH and provide a promising avenue for testing general relativity with strong and dynamical gravitational fields, including the `no-hair' theorem \cite{Berti_2009, PhysRevLett.26.331, Dreyer_2004, berti2025blackholespectroscopytheory}.

The ground-based GW observatories LIGO \cite{LIGO2015} and Virgo \cite{VIRGO} have now detected several tens of high-mass binary BH mergers \cite{2025arXiv250818082T}, with many more expected in the upcoming observing run \cite{Abbott_2020}. 
The fundamental QNM is confidently identified in many of these events \cite{2021arXiv211206861T, 2025arXiv250907348T, 2025arXiv250908099T}, with signs of additional QNMs \cite{Capano_2023}, possibly including overtones \cite{Cotesta_2022, Finch_2022, correia2024lowevidenceringdownovertone}, in several. 
These observations have already been used to test the BH no-hair \cite{Isi_2019} and area theorems \cite{Cabero_2018, Isi_2021, 2025PhRvL.135k1403A}. 

BH ringdown can also be studied using numerical relativity (NR) by fitting QNM models to the ringdown waveform. This can be done either by analyzing the GW signal extracted in a particular viewing direction, by fitting to individual spherical harmonic modes (it is common to look at just the $(2,2)$ spherical mode, for example, see \cite{Giesler_2019}), or by fitting multiple modes simultaneously \cite{Cook_2020,2021arXiv211015922M,zertuche2025multimoderingdownmodellingtextttqnmfits,2014PhRvD..90l4032L,2018arXiv180108208L,2024PhRvD.109d4069C,2024PhRvD.110j3037P} and effectively averaging over viewing directions. 
Traditionally, this is done using linear least-squares fitting (for example, see \cite{Giesler_2019, 2025PhRvD.111b4002D, Mitman_2023}), treating the QNM amplitudes as free parameters. Other techniques, such as nonlinear extensions to least-squares, have also been proposed.
However, these approaches face known challenges, including ambiguity in selecting the appropriate QNM content and the risk of overfitting \cite{zertuche2025multimoderingdownmodellingtextttqnmfits, Mitman_2023, Giesler_2019, 2023PhRvL.130h1401C, 2024CQGra..41s5023T, 2025PhRvD.112h4076V}. 
Moreover, least-squares fitting implicitly assumes stationary white noise and yields only point estimates (without uncertainties) for the QNM model parameters. Another alternative are rational filters \cite{2022PhRvD.106h4036M}, however while these can identify QNMs, they do not give their amplitudes.

As NR simulations and waveform extraction techniques improve, and as GW catalogs expand, there is a growing demand for accurate and physically motivated approaches to QNM modeling. 
A natural progression is a Bayesian approach, which accounts for fit uncertainties via a posterior distribution over the QNM parameters.
There have already been a few attempts at performing Bayesian QNM fits to NR \cite{Clarke_2024, 2024PhRvD.109j1503R, 2024JCAP...10..061C} and this paper introduces a new framework which extends this in several ways.
We introduce a new physically-motivated Gaussian process (GP) model which is trained on an NR catalog.
Additionally, the QNM model is expressed in a way which (with a natural choice of prior) makes the posterior a multivariate normal distribution which can be efficiently sampled. This removes the need for Markov chain Monte Carlo methods which do not scale well to models involving many QNMs. This makes the computational cost of the new Bayesian approach comparable to that of least-squares fitting (typically $\mathcal{O}(1\, \rm ms)$ for a single-mode ringdown fit), therefore making it a viable alternative and supports scalable analyses across large catalogs in the future. 

The Bayesian approach naturally yields posterior distributions over the QNM parameters, as opposed to just point estimates obtained from least-squares fitting; this is a key advantage of the new Bayesian approach.
The posterior provides the foundation for assigning a measure of significance to the various QNMs in the model and for performing a selection between models using different numbers or types of QNMs \cite{content_paper}. 
This is advantageous because point estimates alone cannot determine whether or not the NR data supports including a specific QNM in the model.
This is expected to be increasingly important in future studies focusing on subdominant and higher order effects in the ringdown such as nonlinear quadratic QNMs \cite{Mitman_2023, 2023PhRvL.130h1401C} and power-law tails \cite{2024arXiv241206906M}.

This paper uses the public catalog of 13 Cauchy-Characteristic Evolved (CCE) waveforms produced by the Spectral Einstein Code (SpEC) \cite{spectrecode, Moxon2023}. 
These waveforms more accurately extract the GW information at future null infinity than extrapolated methods \cite{Moxon2023}. 
This information is contained in the Bondi news function $\mathcal{N}$ (although the strain $h$ or the Weyl curvature scalar $\Psi_4$ can also be used). 

A GP is used to model the numerical uncertainties in these waveforms. The model is trained on the full catalog of CCE simulation residuals to create a bespoke kernel which is used to define the Bayesian likelihood. This provides a physically-motivated, flexible, and conservative estimate of the waveform uncertainty in the catalog.

This is a methods paper, in which the Bayesian QNM inference procedure and GP are described. This new approach is demonstrated by applying the techniques to two simple models. Results from these analyses showcase the potential of this methodology and provide a case study for the significance and posterior predictive check which are also introduced.  

%%%
\section{QNM Model and Likelihood Function} \label{sec:likelihood}

The model described in this section can be applied to any of the GW quantities $h$, $\mathcal{N}$, or $\Psi_4$. For concreteness, all quantities in this section are written in terms of $h$. 

The complex GW strain, $h=h_+-ih_\times$ can be expanded in (spin-weight) spherical harmonics, \sYlm{-2}{\ell}{m} as
\begin{align}\label{eq:Ylm_expansion}
    r\mathfrak{h}(t,\theta,\phi) = M \sum_{\beta} \mathfrak{h}^{\beta}(t) ~ {}_{-2}Y^\beta(\theta, \phi) ,
\end{align}
where $\beta=(\ell,m)$ denotes the pair of spherical harmonic indices and the sum extends over $\ell\geq2$ and $|m|\leq \ell$.
The harmonic modes $\mathfrak{h}^{\beta}(t)$ are obtained as an output of NR simulations (see Sec.~\ref{sec:CCEcatalog}).

The QNM model for the GW strain consists of a sum of damped sinusoids. Following the notation of Ref.~\cite{2025PhRvD.111b4002D}, this model is written as
\begin{align}\label{eq:QNMmodel}
    rh(t,\theta,\phi)  = 
    M \sum_{\alpha} C_{\alpha} e^{-i \omega_{\alpha}(t-t_0)} {}_{-2}S_{\alpha}(\chi; \theta, \phi),
\end{align}
where $\alpha\!=\!(\ell,m,n,p)$ denotes the quadruple of spheroidal harmonic indices that uniquely identifies a QNM and the sum extends over $\ell\geq2$, $|m|\leq \ell$, $n\geq 0$, and $p\in\{-,+\}$. 
The quantity $t_0$ is the ringdown start time and the model in Eq.~\ref{eq:QNMmodel} is understood to apply for $t>t_0$.
In practice, it is also necessary to choose an end time $T$ for the analysis, so $t_0\leq t\leq T + t_0$.
This QNM model has many free amplitude parameters $C_\alpha$.
Optionally, the remnant BH's mass and spin, $M_f$ and $\chi_f$, can be included as free parameters. 
The start time $t_0$ is treated as fixed.

When the model described in Eq.~\ref{eq:QNMmodel} is fit to the strain, the resulting amplitudes are given in the so-called `strain-domain'. The Bondi news, which is the first derivative with respect to time of the strain, can also be fit using this model with the amplitudes in the `news-domain'. 
The same applies to the Weyl curvature scalar, which is the second derivative with respect to time of the strain. 
The various amplitudes are related as follows,
\begin{align} \label{eq:domains}
    C_{\alpha}[h] = iC_{\alpha}[\mathcal{N}] / \omega_{\alpha} = -C_{\alpha}[\Psi_4]/\omega_a^2, 
\end{align}
where square brackets have been used to indicate the respective domain of the amplitude. 
The units of $C_{\alpha}[h]$, $C_{\alpha}[\mathcal{N}]$, and $C_{\alpha}[\Psi_4]$ are 1, $M^{-1}$, and $M^{-2}$ respectively.

The spherical and spheroidal harmonics are related by the mode-mixing coefficients $\mu^{\beta}_{\alpha}$ (see, for example, Eq.~4 of Ref.~\cite{2025PhRvD.111b4002D}) which allows the QNM model in Eq.~\ref{eq:QNMmodel} to be re-expanded in the spherical harmonic basis; 
\begin{align} \label{eq:QNM_model_modes}
    rh(t,\theta,\phi)  &= \sum_\beta rh^\beta(t) ~ {}_{-2}Y^\beta(\theta, \phi) , \\ 
    \mathrm{where} \quad r h^\beta (t) &= M \sum_\alpha \mu_\alpha^\beta C_\alpha e^{-i \omega_{\alpha}(t-t_0)} \nonumber .
\end{align}
The $\mu^{\beta}_{\alpha}$ coefficients and $\omega_\alpha$ frequencies were calculated as functions of $M_f$ and $\chi_f$ using the \texttt{qnm} package \cite{Stein:2019mop}.

Both the NR data and QNM model are discretely sampled time series evaluated at sample times $t_a$.
Indices $a,b,\ldots\in\{1,2,\ldots N\}$ label points in the time series.
If the numerical uncertainty on the waveform modes $\mathfrak{h}^\beta$ is assumed to be independent and distributed as a zero-mean Gaussian in each spherical harmonic mode, then the log-likelihood function is given by
\begin{align} \label{eq:loglike}
    \log P(\mathfrak{h}|\boldsymbol{\theta}) = \sum_\beta \frac{-1}{2} \left<\mathfrak{h}^\beta-h^\beta|\mathfrak{h}^\beta-h^\beta\right>_\beta ,
\end{align}
where the data are the NR waveform modes (denoted collectively by $\mathfrak{h}$) and the model parameters $\boldsymbol{\theta}$ are the complex mode amplitudes $C_\alpha$ and (possibly) $M_f$ and $\chi_f$.

In Eq.~\ref{eq:loglike}, the angle brackets with a subscript $\beta$ denote an inner product between two functions of time, $f(t)$ and $g(t)$ (for each spherical $\beta$ mode) discretely sampled at the times $t_a$. This inner product is given by
\begin{align}
    \left<f|g\right>_\beta = \mathrm{Re}\,\left\{\sum_{a,b} f^*_a g^{\phantom{*}}_{b} \left(\left(\mathbf{K}^{\beta}\right)^{-1}\right)_{ab} \right\},
    \label{eq:innerproduct}
\end{align}
where $\mathbf{K}^{\beta}$ is the covariance matrix in each mode and a star $^*$ denotes complex conjugation. 
This covariance is designed to model the numerical uncertainty in each spherical waveform mode.
This is modeled using a GP that is trained on a public catalog of waveforms that are described in Secs.~\ref{sec:CCEcatalog} and \ref{sec:GPcov}.

The inner product in Eq.~\ref{eq:innerproduct} explicitly depends on the choice of sample times, $t_a$; these are usually chosen to be regularly spaced in the interval $[t_0, T + t_0]$ with some sampling frequency $1/\Delta t$. As the sampling frequency is increased and the sample points become dense in $[t_0, T + t_0]$ this sum converges to the reproducing kernel Hilbert space (RKHS) $\mathcal{H}_k$ inner product $\left<f|g\right>_{\mathcal{H}_k}$ \cite{Rasmussen2006Gaussian}. This is verified in Appendix \ref{app:sampling_frequency}. 

If the components of all the covariance matrices $\mathbf{K}_\beta$ are taken to be $\sigma^2 \delta_{ij}$ (i.e.\ white noise, with the same $\sigma$ for each $\beta$ mode) and the maximum likelihood value of $\boldsymbol{\theta}$ is used as a point estimator for the parameters of the QNM model, then this reduces to the least-squares minimization methods that have been used almost exclusively so far in the literature.
This paper improves on this in two ways. 
Firstly, instead of using a white-noise covariance, we introduce a physically-motivated GP model for the uncertainty in the NR waveforms and numerically train this model on a NR waveform catalog.
Secondly, instead of using a point estimate for the QNM model parameters, we place a prior on $\boldsymbol{\theta}$ and analytically sample from the resulting posterior.

Expressing the complex mode amplitudes in terms of their real and imaginary parts, $C_\alpha=\mathrm{Re}\,C_\alpha{}+i\mathrm{Im}\,C_\alpha{}$, and treating these as the ($2N$) parameters, the QNM model in Eq.~\ref{eq:QNMmodel} is fully linear. 
Additionally, if the remnant BH mass and spin, $M_f$ and $\chi_f$, are to be included as free parameters then we linearize the model by Taylor expanding about a chosen reference point. 
This reference point is taken by performing a linear least-squares fit of the QNM model to the NR data, using the fixed asymptotic Bondi data (ABD) values of the mass and spin. The best fitting values for the amplitudes, in addition to the ABD mass and spin, are taken as the reference parameters and are denoted $\boldsymbol{\theta}_*$. The best fitting model evaluated at these parameters is denoted $H^\beta_*(t)$. Alternatively, a non-linear least-squares fit can be used to obtain mass and spin reference values, in addition to the amplitudes. This method is slower and does not significantly improve the fits. This is discussed further in App. \ref{app:linearisedapproximation}. 

The QNM model in Eq.~\ref{eq:QNMmodel} is expanded to first order in changes in the model parameters about the chosen reference. In other words, the model is put into the form
\begin{align} \label{eq:simple_model_linearised}
    h^\beta(t) = H_*^\beta(t) + \sum_\mu\left(\boldsymbol{\theta}^\mu -\boldsymbol{\theta}^\mu_*\right)h^\beta_\mu(t) .
\end{align}
The quantities $h^\beta_\mu(t)$, sometimes called the model matrix, are built from the first derivatives of the model in Eq.~\ref{eq:QNMmodel} evaluated at the reference. Most of the differentiation involved in constructing the model matrix can be done analytically. The exceptions are the derivatives of the QNM frequencies $\omega_\alpha$ and the mode-mixing coefficients $\mu_\alpha^\beta$ with respect the remnant BH spin $\chi_f$; these derivatives were taken numerically using finite differences.

The linearization of the QNM model in the mass and spin parameters necessarily introduces an additional source of error. 
However, this is small because typically fitting to state-of-the-art NR waveforms leads to small uncertainties in these parameters. The typical maximum fractional error introduced in the QNM frequencies is less than $\sim 10^{-7}$.
Quantities like $\omega_\alpha$ and $\mu_\alpha^\beta$ that depend on these  parameters vary only slightly across the width of the typical set of the posterior and this is expected to be well approximated by a linear expansion about a suitably chosen reference point. 
The accuracy of the linear approximation is discussed further below and is quantified in App.~\ref{app:linearisedapproximation}. 

Initially, flat priors were used on all model parameters, $\pi(\boldsymbol{\theta})=\mathrm{const}$. 
Hereafter, this is referred to as `Prior 1'.
(It would also be possible to use a multivariate Gaussian prior on $\boldsymbol{\theta}$ without significantly changing any of the following conclusions.) Because the model is linear in all of its parameters, with this choice of prior the posterior is exactly a multivariate Gaussian;
\begin{align} \label{eq:Gaussian_posterior}
    \boldsymbol{\theta}|\mathfrak{h} &\sim \mathcal{N}\big(m^\mu, (\Gamma^{-1})^{\mu\nu}\big) , \\
    \mathrm{where} \quad m^\mu &= \boldsymbol{\theta}^\mu_* + (\Gamma^{-1})^{\mu\rho}\sum_\beta\left<h^\beta_\rho|d^\beta-H^\beta_*\right>_\beta , \nonumber\\
    \mathrm{and} \quad
    \Gamma_{\mu\nu} &= \sum_\beta \left<h^\beta_\mu|h^\beta_\nu\right>_\beta. \nonumber
\end{align}
This multivariate Gaussian posterior can be sampled extremely efficiently.
First, the posterior covariance $\Gamma_{\mu\nu}$ (numerically equal to the Fisher information matrix) is evaluated and inverted using its eigenvalue/vector decomposition (computed using \texttt{scipy.linalg.eigh}).
For reasons of numerical stability, it is necessary to regularize some of the very small eigenvalues of the Fisher matrix; any eigenvalues smaller than a specified tolerance value $\epsilon=10^{-10}$ were replaced with the value $\epsilon$.
The mean vector $m^\mu$ and covariance matrix $\Gamma^{-1}$ were then calculated and the posterior in Eq.~\ref{eq:Gaussian_posterior} is then sampled analytically.

The ability to sample the multivariate Gaussian posterior in Eq.~\ref{eq:Gaussian_posterior} analytically is key to the efficiency of the method. 
Moreover, the posterior only has a Gaussian form because of the choice of prior described above: a flat prior on the real and imaginary parts of the amplitudes.
It is sometimes desirable to switch to using a prior that is instead flat on the mode amplitudes $|C_\alpha|$ and flat and periodic in the phase angles $\mathrm{arg}(C_\alpha)$. 
Hereafter, this choice is referred to as `Prior 2'. 
Sampling directly with this prior would be much less efficient because the posterior would no longer be a multivariate Gaussian.
Therefore, if samples with Prior-2 posterior are needed, these are obtained by first sampling the  Prior-1 posterior, as described above, and then using importance sampling to reweight the samples to the new prior.
The importance sampling weights are given by
\begin{align} \label{eq:importance_weights}
    w_n = \prod_\alpha \frac{1}{|C_{\alpha,n}|},\; \mathrm{for}\;n=1,2,\ldots,N_{\rm samples},
\end{align}
where $n$ indexes the posterior samples.

Care must be taken when using importance sampling that the effective number of independent samples doesn't drop too low. 
In the present case, this is liable to happen when there are a large number of QNMs included in the model with small amplitudes (see Eq.~\ref{eq:importance_weights}). 
The effective number of independent samples is quantified using \cite{Kong1992}
\begin{align}
    n_{\rm eff} = \frac{\left(\sum_{n}w_n\right)^2}{\sum_{n} w_n^2} \leq N_{\rm samples} .
\label{eq:neff}
\end{align}

A major benefit of sampling from the full Bayesian posterior on the QNM mode parameters, as compared with earlier approaches that relied on point estimates, is that the extent of the posterior can be used to quantify the support for the point $C_\alpha=0$ for any given mode $\alpha$. This point corresponds to that QNM not being included in the model.
This allows us to assign a well-motivated Bayesian measure of significance $\mathcal{S}_\alpha$ to each QNM in the fit.
This significance is closely related to the Bayes factor, or evidence ratio, between two QNM models with and without the $\alpha$ QNM included (but otherwise with identical QNM content). 
The precise definition of the significance and its relationship to the Bayes factor is explained in App.~\ref{app:significance}.

%%%
\section{CCE Waveform Catalog} \label{sec:CCEcatalog}

The waveforms used in this study come from the public catalog \cite{SXS_CCE_catalog} of 13 Spectral Einstein Code (SpEC) waveforms \cite{Boyle:2019kee} extracted using SpECTRE CCE \cite{spectrecode, Moxon_2020, Moxon2023}.
The simulations were transformed into the superrest frame $300\,M$ after the peak strain using \texttt{scri} \cite{mike_boyle_2020_4041972, mike_boyle_2020_4041972, Boyle2013, BoyleEtAl:2014, Boyle2015a} and shifted in time so that the peak of the $\ell=m=2$ mode strain occurs at $t=0$ \cite{2024CQGra..41v3001M, Mitman:2022kwt, 2021arXiv211015922M}.

In principle, the methods described here could be applied to either the strain $\mathfrak{h}$, the news $\mathcal{N}$, or the Weyl curvature scalar $\psi_4$ (see App.~\ref{app:full_param}). 
Typically, fits are done to the strain, however after performing the supertranslation on the simulations, there can be some residual error which manifests as the strain not perfectly decaying to zero. These imperfections are not present in the news, which always decays to zero \cite{2025arXiv250309678M}. Further details of this are discussed in App.~\ref{app:residual_comparison}. 
Therefore in this study we choose to work with the news and report amplitudes in the news-domain.

For each simulation, the highest resolution level (L5) extracted at the second smallest worldtube radius was used as the preferred waveform.

Information about the uncertainty in the NR waveform can be found in the residuals $\mathfrak{r}^\beta(t)$, defined as the difference from the second-highest resolution level (L4) at the same worldtube radius. This gives an estimate of the uncertainty on the second-highest resolution level, however here it is used as a \emph{conservative} estimate for the uncertainty on the highest resolution waveform. Although crude, this approach to estimating waveform uncertainties is relatively standard in the field of numerical relativity. In principle, more sophisticated probabilistic numerical methods \cite{HenOsbGirRSPA2015} may one day provide an improved description of the numerical uncertainty.
Full details on the processing of the numerical waveforms are given in App.~\ref{app:data_processing}. 

These waveform residuals are used to train a GP model for the uncertainty in the preferred waveform; this is described in Sec.~\ref{sec:GPcov}.

%%%
\section{GP Model for the Waveform Uncertainties} \label{sec:GPcov}

\begin{figure}[t]
    \centering
    \includegraphics[width=0.48\textwidth]{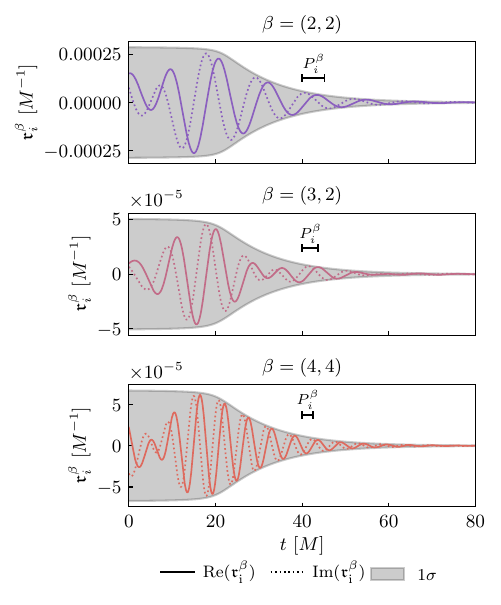}
    \caption{ \label{fig:gp_shaded_regions}
        These plots illustrate the GP model (Eq.~\ref{eq:kernel}) for the uncertainty in selected waveform modes ($\beta=$(2,2), (3,2), and (4,4) in the top middle and bottom panels respectively) of simulation $i=0001$. 
        The colored curves show the real and imaginary parts of the residuals in the Bondi news, $\mathfrak{r}_i^\beta(t)$.
        The shaded regions show the amplitude of GP model for the waveform uncertainties.
        What cannot be seen from the shaded regions is the timescale over which the GP model is correlated;
        this is illustrated with the horizontal bars on each panel, scaled to the squared-exponential timescale as described in the text.
    }
\end{figure}

The purpose of this section is to describe the model for the uncertainty in NR merger and ringdown waveforms. 
This is used to define the QNM likelihood described in Sec.~\ref{sec:likelihood}.
The model describes the uncertainty in the real and imaginary parts of the waveform in each spherical harmonic mode $\beta$ in each simulation $i$ of the catalog independently.
The model is taken to be a GP with a physically-motivated choice for the kernel that includes a number of free parameters. 
The values of these free parameters are then learned by maximizing the likelihood of the GP model evaluated on the training data which is taken to be the residuals $\mathfrak{r}_i^\beta(t)$ for each waveform mode of each simulation in the catalog. 
The resulting model for the uncertainty is illustrated in Fig.~\ref{fig:gp_shaded_regions} where it is plotted alongside some of the residuals that were used in the training.
The rest of this section describes the details of this GP model.

Examination of the residuals shows that they share many properties with the waveforms themselves. The functions $\mathfrak{r}_i^{\beta}(t)$ are smooth and vary on timescales comparable to the GW period. The amplitude of the residuals also tracks the relevant quantity (either the strain, news, or curvature scalar), with the largest residuals found in the $(\ell, m) = (2, 2)$ mode around merger. The residuals also show common behavior across all the simulations in the catalog. This is exploited here by pooling kernel GP parameters to model the waveform uncertainty, with kernel properties inferred by training the GP on the full catalog, resulting in a set of GP parameters which are learned from the waveforms used. 

After experimenting with a number of options, the following kernel, hereafter the \emph{standard kernel} (or simply GP for short), was chosen to model the numerical uncertainty in the waveforms, 
\begin{align}\label{eq:kernel}
&k_{i}^{\beta}(t, t') 
    = k_{\rm Stationary}(t, t'; P_{i}^{\beta}) \,  \\
    &\quad\times \operatorname{\textsc{SmoothMin}}\!\left(
         \sigma_{0,i}^\beta e^{-t/\tau_{i}^{\beta}},\ 
         \sigma_{\mathrm{max},i}^\beta;\ s
       \right) \nonumber \\
  &\quad\times \operatorname{\textsc{SmoothMin}}\!\left(
         \sigma_{0,i}^\beta e^{-t'/\tau_{i}^{\beta}},\ 
         \sigma_{\mathrm{max},i}^\beta;\ s
       \right). \nonumber
\end{align}
The first term in Eq.~\ref{eq:kernel} is a stationary kernel, which is intended to capture the fact that the residuals vary on a predictable timescale, $P_{i}^{\beta}$ related to the real part of the fundamental QNM frequency. 

One possible choice for $k_{\rm Stationary}(t, t'; P_{i}^{\beta})$ would be the squared-exponential (SE) kernel, which would capture the periodicity of the residual. 
However, this results in a dense matrix with entries that become very small compared to the diagonal values at large $|t - t'|$. 
This can slow computations and increase numerical instability. 
For this reason, we instead use a $q = 3$ Wendland covariance function \cite{Rasmussen2006Gaussian} (see App.~\ref{app:wendland} for details). 
This mimics the SE kernel but with compact support leading to a band structure in the covariance matrix.
We find this choice improves numerical stability when inverting the covariance matrix. 
In principle the sparse structure of the matrix can also speed up computations when using the appropriate solver functions, however we do not take advantage of this here. 

The timescale $P_{i}^{\beta}$ is closely related to period of the GW oscillations and is therefore expected to be close to the real part of the frequency $\omega_{\alpha}$ of the fundamental ($n=0$) prograde QNM with the same $\ell$ and $m$. 
Therefore, this timescale is taken to be
\begin{align}
    P_{i}^{\beta} = \frac{2\pi\mu}{\mathrm{Re}\{\omega_{\ell m 0 +}(M_{f,i},\chi_{f,i})\}}.
\end{align}
The QNM frequencies are functions of the remnant mass and spin, $M_{f,i}$ and $\chi_{f,i}$, for each simulation $i$ in the catalog which are known from the NR simulation; specifically, these values were obtained from the ABD object in the \texttt{scri} package.
The pooled GP parameter $\mu$ is introduced to adjust this estimate for the timescale and is pooled across all spherical harmonic modes in all simulations in the catalog. 

The timescale in the Wendland covariance function can be related to the timescale in the squared-exponential function by equating their second derivatives at the origin. This gives $\mathrm{SE  \,\, Timescale} = \sqrt{5/72}\,P_i^{\beta}$. This timescale is plotted using the horizontal bars in Fig.~\ref{fig:gp_shaded_regions}. 

\begin{figure}[t]
    \centering
    \includegraphics[width=0.48\textwidth]{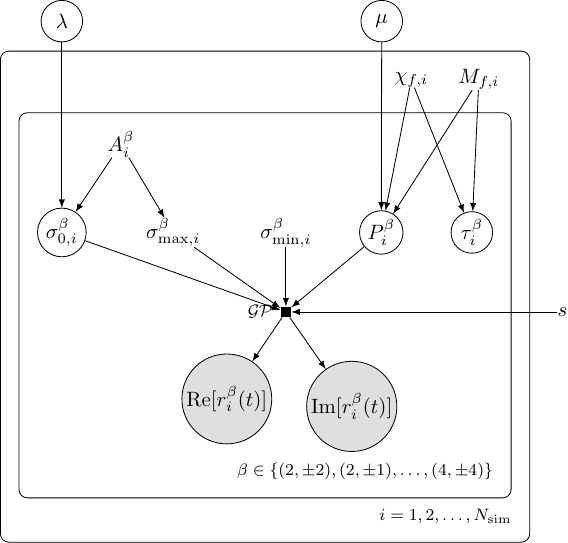}
    \caption{ \label{fig:pgm}
        PGM of the GP model for the NR waveform uncertainty described in the main text. 
        The observed variables (shaded circles) are the real and imaginary parts of the waveform residuals in each simulation in the catalog (indexed by $i$) and in each spherical harmonic model (indexed by $\beta$).
        The residuals are the differences between the two highest resolution simulation available in the catalog.
        The parameters of the model are shown in empty circles. 
        The values of the two pooled parameters in the top row are ultimately to be inferred from the NR catalog.
        The other latent parameters are determined from the pooled parameters and from the various fixed quantities which are shown without circles.
    }
\end{figure}

The next terms in Eq.~\ref{eq:kernel} control the time dependence of the typical amplitude of the residuals. 
The function, 
\begin{align}
    \sigma_{0,i}^\beta \exp(-\frac{t}{\tau_{i}^{\beta}}),
\end{align}
imposes the expected decay during the ringdown. 
The scale $\sigma_{0,i}^\beta$ controls the amplitude of the residuals and is estimated using the maximum amplitude $A^\beta_i$ of the numerical residuals, 
\begin{align}
    \sigma_{0,i}^\beta = \lambda A^\beta_i,\quad\mathrm{where}\quad A^\beta_i=\mathrm{max}_t\, |\mathfrak{r}_i^{\beta}(t)|.
\end{align}
The function $\mathrm{max}_t$ determines the maximum of the residuals over the full range of times, therefore the constants $\sigma_{0,i}^\beta$ and $A^\beta_i$ themselves do not depend on time. The pooled GP parameter $\lambda$ tunes this estimate. 

The parameter $\tau_{i}^{\beta}$ controls the rate of the exponential decay and is estimated using the imaginary part of the fundamental prograde QNM with the same $\ell$ and $m$,
\begin{align}
    \tau_{i}^{\beta} = \frac{-1}{\mathrm{Im}\{\omega_{\ell m 0 +}(M_{f,i},\chi_{f,i})\}}.
\end{align}

The exponential decay only occurs during the ringdown. During the merger, the amplitude of the residuals is expected to be roughly constant with time. This is achieved in our model using the \textsc{SmoothMin} function which limits the maximum value of output to 
\begin{align}
{\sigma_{\mathrm{max},i}^\beta = \nu A^{\beta}_{i}}
\end{align}
at early times, where we fix $\nu = 1.1$. At intermediate times, the function smoothly transitions to the exponential decay with the smoothness controlled by the parameter $s$ which was fixed to $10^{-3}$. Further details about the \textsc{SmoothMin} function are given in App.~\ref{app:smoothmin}.

At very late times, the residuals are expected to, again, become roughly constant with time when the signal has decayed below the level of the numerical noise floor of the simulation. 
To model this, and to prevent issues relating to vanishing numerical uncertainties and ensure numerical stability we add a small jitter term along the diagonal of covariance matrix that is scaled relative to the average amplitude of the late-time residuals $\sigma_{\mathrm{min},i}^{\beta}$, 
\begin{align}
    &\sigma_{\mathrm{min},i}^{\beta}
        = \frac{1}{t_{2} - t_{1}} \int_{t_{1}}^{t_{2}} \left| \mathfrak{r}_i^{\beta}(t) \right| \, dt \\
    &k_{\rm Jitter}(t_a, t_b) = \big(\epsilon \, \sigma_{\mathrm{min},i}^\beta\big)^2 \, \delta_{ab},
\end{align}
where values of $t_2 = 300$ and $t_1 = 250$ provided a sufficient range over which to approximate the late-time constant. We fix $\epsilon = 1 \times 10^{2}$, which prevents any eigenvalues of the covariance matrix becoming negative and gives a conservative estimate of the late-time uncertainty. The jitter cannot be seen in Fig.~\ref{fig:gp_shaded_regions} because it is too small. However, it can be seen clearly in Fig.~\ref{fig:gp_shaded_regions_logarithmic} in App.~\ref{app:reg_gp}. 

\begin{table}[t]
\caption{\label{tab:modes}
    The spherical harmonic modes for each simulation that were used for training the GP model. 
    We focus on the loudest modes $\beta=(2,\pm 2),\,(2,\pm 1),\,(3,\pm 3),\,(3,\pm 2),\,(4,\pm 4)$ with some omissions on symmetry grounds.
    For the non-precessing simulations 0001-0007 and 0010-0012 the modes satisfy $h_{\ell, m}=h_{\ell, -m}^*$, hence the $m<0$ modes are excluded as they carry no extra information. 
    For the equal-mass, non-precessing, identical-spin simulations 0001-0004 (and the `superkick' simulation 0009) the odd $m$ modes are suppressed and were excluded. 
}
\begin{ruledtabular}
\begin{tabular}{lr}
    Simulation ID & GP Training Modes \\
    \hline
    0001-0004 & $(2,2), (3,2), (4,4)$ \\
    0005-0007, 0010-0012 & $(2,2), (2,1), (3,3), (3,2), (4,4)$  \\
    0008, 0013 & $(2,\pm 2), (2,\pm 1), (3,\pm 2), (3,\pm 3), (4,\pm 4)$ \\
    0009 & $(2,\pm 2), (3,\pm 2), (4,\pm4)$  \\
\end{tabular}
\end{ruledtabular}
\end{table}

In total, the standard GP kernel, for a spherical harmonic mode of a specific NR simulation, is determined by six parameters: $\sigma^\beta_{0,i}$, $\sigma_{\mathrm{max},i}$, $\sigma_{\mathrm{min},i}$, $P^\beta_i$, $\tau^\beta_i$, and $s$. An $i$ index indicates a parameter that depends on the NR simulation being studied, while a $\beta$ index indicates a parameter that depends on the spherical harmonic mode.
These parameters are in turn either fixed or controlled by the pooled GP parameters $\lambda$ and $\mu$ which are pooled across all modes of all simulations. 
The kernel also depends on several quantities that are taken directly from the highest resolution (L5) NR simulation: $M_{f,i}$, $\chi_{f,i}$, or the residual of the L5 and L4 simulations: $A^\beta_i$.
The dependencies of all the parameters in this model for the waveform uncertainty are illustrated in the probabilistic graphical model (PGM) in Fig.~\ref{fig:pgm}.

The kernel is used to build the covariance matrix, $\mathbf{K}^\beta$, for each simulation $i$ by evaluating the kernel at all pairs of times in the waveform time series; 
\begin{align}
    \left(\mathbf{K}_{i}^{\beta}\right)_{ab} = k_i^\beta(t_a, t_b).
\end{align}

The pooled GP kernel parameters $\boldsymbol{\psi} = \{ \lambda, \mu \}$ are learned from the public catalog of $N_{\rm sim}=13$ CCE simulations by maximizing the GP log-likelihood across all the waveform modes in all of these simulations. 
The real and imaginary parts of the uncertainty on each mode of each simulation are assumed to be independent.
The GP log-likelihood is \cite{Rasmussen2006Gaussian}
\begin{align} \label{eq:logz}
    \log P\left( \big\{ \mathfrak{r}_i^{\beta} \big\} | \boldsymbol{\psi}\right) \!=\! \dfrac{-1}{2} \! \sum_{{\rm X},i,\beta}\!&\Big[({\rm X} \, \mathfrak{r}_i^{\beta})^{\rm T}(\mathbf{K}_{i}^{\beta})^{-1}({\rm X} \, \mathfrak{r}_i^{\beta}) \\
    &\!+\! \log(|\mathbf{K}_{i}^{\beta}|) \!+\! n\log(2\pi) \Big], \nonumber
\end{align}
where ${\rm X}\in\{\mathrm{Re}, \mathrm{Im}\}$ denotes the real or imaginary part and $i=1,\ldots, N_{\rm sim}$.

In practice, when training the GP using Eq.~\ref{eq:logz}, only modes $\beta\in\{(2,\pm 2), (2,\pm 1), (3,\pm 3), (3,\pm 2), (4,\pm 4)\}$ were used. These were chosen because they are typically expected to be the loudest. Using this restricted set of modes reduces the cost of evaluating Eq.~\ref{eq:logz}, thereby speeding up the training.
We also deliberately exclude $m=0$ modes which have previously been shown to behave differently (e.g. see \cite{2025PhRvD.111b4002D}) and we suspect may need to be treated independently from the $m \neq 0$ modes when determining their numerical uncertainty contribution. 
For systems with certain symmetries the number of modes was further restricted: e.g.\ equal-mass, non-precessing, identical-spin, only the even $m$ modes were included. 
For details of the exact modes used, see Table \ref{tab:modes}.  
Although the model is only trained on modes with $\ell\leq4$, the GP model nevertheless provides a good description of the numerical uncertainties in higher-order modes (with $m \neq 0)$ as well. This is demonstrated up to $\ell\leq 7$ in Fig.~\ref{fig:gp_shaded_regions_logarithmic} of App.~\ref{app:reg_gp}.

The negative log-likelihood was minimized using \texttt{scipy.optimize.minimize} employing the Nelder-Mead method. (Maximizing the likelihood implicitly assumes flat priors on all of the $\boldsymbol{\psi}$ parameters in Eq.~\ref{eq:hyperparam_values}.) As the GP training is a one-off exercise, we did not attempt to optimize the training cost beyond the standard \texttt{scipy} minimization. On a standard desktop computer, the training took $\sim 5$ minutes for the GP kernel. For the simple white-noise kernel and `complicated kernel' described in the following section, the training took $\sim 8$ minutes and $\sim 25$ minutes respectively. The cost of evaluating the likelihood function scales linearly with the number of waveforms in the catalog and therefore this method should scale well to larger catalogs. However, although these costs are low, they can depend sensitively on the range of parameters explored and the details of the GP likelihood function and can sometimes be difficult to optimize without a good initial guess of the GP parameters.

In the GP case, the model was trained on the residual time spanning $20M\leq t\leq 80M$ during which the model can efficiently tune to the periodicity and amplitude of the residual during ringdown. In the WN case, the residual was trained on the range $0\leq t\leq 200M$ so as to also account for the early- and late-time data. 
The pooled GP parameters were allowed to vary in the following ranges: $\lambda\in [0.01, 50]$ and $\mu \in [0.1, 10]$. 
This training yielded the following values for the GP parameters:
\begin{align} \label{eq:hyperparam_values}
    \boldsymbol{\psi}_{\rm GP}=(\lambda, \mu) = (6.92, 1.68). 
\end{align}

We note here that the GP model for the waveform uncertainty and the resulting posteriors on the QNM parameters are not sensitive to the choice of the sampling frequency. This is demonstrated in App.~\ref{app:sampling_frequency}.

%%%
\section{Comparison of GP uncertainty models} \label{sec:GPresults}

The standard (GP) kernel described in Sec.~\ref{sec:GPcov} provides a well-motivated description of the NR waveform noise, with parameters that are based on physically-relevant features in the ringdown. However, many choices went into the design of this model and these are clearly far from being unique. In order to validate the model and to demonstrate both that it can not be significantly improved upon by introducing additional flexibility with more free pooling parameters and, conversely, that it is not already too flexible and is overfitting the training data, it was compared to several other kernel models. This section presents the results comparing the GP model described above to two alternatives: one simpler model (the white noise kernel, or WN) and one more complicated model (the `complicated' kernel, or GPc).

First, the simple, white-noise kernel function is defined as
\begin{align} \label{eq:kernel_white_noise}
    k^\beta_i(t_a,t_b) = \sigma_{0,i}^\beta \,\delta_{ab} , \quad \mathrm{where} \quad \sigma_{0,i}^\beta = \lambda A^\beta_i ,
\end{align}
where $\lambda$ is the sole pooled parameter that, after training, takes the value
\begin{align}
    \boldsymbol{\psi}_{\rm WN}=(\lambda) = (0.20).
\end{align}
This kernel model is extremely simple with just a single pooled parameter (the PGM for this model is not shown). 
If Fig.~\ref{fig:gp_shaded_regions} were replotted using this simple WN kernel then the width of the shaded regions would be large and constant for all times.

In the unique case of a simple white noise kernel, which has a diagonal covariance matrix, the inner product defined in Eq. \ref{eq:innerproduct} must be modified by multiplying by a factor of the time step $\Delta t$. This accounts for scaling that occurs as the density of times used to construct the matrix is changed, and ensures the inner product properly converges to an integral in the limit as $\Delta t \rightarrow 0$. 

Another reason for considering the simple WN kernel is that it is closely related to the least-squares fits that have been widely used previously and therefore makes a good comparison.
Specifically, for fixed $M_f$ and $\chi_f$, the maximum \emph{a posteriori} parameter estimator for the QNM parameters $\boldsymbol{\theta}$ obtained using a likelihood defined using covariance matrices $\mathbf{K}^\beta$ built from this simple WN kernel, and using Prior 1, are exactly the same parameters that would be found in the usual least-squares fit.

\begin{figure}[t]
    \includegraphics[width=0.48\textwidth]{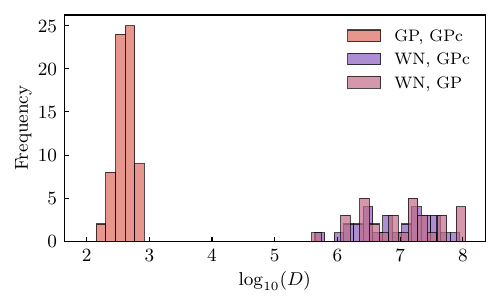}
    \caption{\label{fig:JS_histogram} The symmetrized KLD, $D$, (in bits) between the Gaussian distributions defined using the three kernels across the full catalog and all spherical modes used in training. The difference between the standard kernel (GP) or complicated kernel (GPc) and the simple white noise (WN) kernel are significant, however those between the GP kernel and the GPc are localized at smaller values, indicating that they are more similar.}
\end{figure}

The more complicated kernel function (GPc), uses a flexible mixture of a stationary and periodic kernel. This introduces additional freedom into the kernel that can be used to model the oscillatory part of the residual. The time-dependent model for the amplitude is the same as in the standard (GP) case. The kernel is defined as
\begin{align}\label{eq:kernel_complex}
k_{i}^{\beta}(t,t') 
  &= \big( a\,k_{\mathrm{Stationary}}(t,t';P_{i}^{\beta}) \\
  &\quad+ (1-a)\,k_{\mathrm{Periodic}}(t,t';p^\beta_i,l^\beta_i)\big) \nonumber \\
  &\quad\times \operatorname{\textsc{SmoothMin}}\!\left(
         \sigma_{0,i}^\beta e^{-t/\tau_{i}^{\beta}},\ 
         \sigma_{\mathrm{max},i}^\beta;\ s
       \right) \nonumber \\
  &\quad\times \operatorname{\textsc{SmoothMin}}\!\left(
         \sigma_{0,i}^\beta e^{-t'/\tau_{i}^{\beta}},\ 
         \sigma_{\mathrm{max},i}^\beta;\ s
       \right). \nonumber
\end{align}
where the periodic kernel and its parameters are
\begin{align}
    k_{\rm Periodic}(t,t';p^\beta_i,l^\beta_i) \!=\!  
    \exp\left(\frac{-2\sin^2\left(\frac{\pi|t-t'|}{p^\beta_i}\right)}{(l^\beta_i)^2}\right) , 
\end{align}
where
\begin{align}
    &p^\beta_i = \frac{2\pi\eta}{\mathrm{Re}\{\omega_{\ell m 0 +}(M_{f,i},\chi_{f,i})\}} , \\
    &l^\beta_i = \frac{-\kappa}{\mathrm{Im}\{\omega_{\ell m 0 +}(M_{f,i},\chi_{f,i})\}} , 
\end{align}
and $\eta$ and $\kappa$ are new pooled parameters.
The final pooled parameter $0<a<1$ controls the relative weight given to the squared-exponential and periodic components of the kernel.
Flat priors were used on all three of the new pooled parameters and they were allowed to vary in the following ranges: $\eta\in [0.1, 10]$, $\kappa \in [0.1, 10]$, and $a\in [0, 1]$.

In total, this complicated GPc kernel includes three additional parameters compared to the standard GP kernel described above: $a$, $\eta$, and $\kappa$. 
The structure of this more complicated kernel model is illustrated in the PGM shown in App.~\ref{app:GPC}.
Training this GPc model yielded the following parameter values,
\begin{align}
    \boldsymbol{\psi}_{\rm GPc}=(\lambda, \mu, \eta, \kappa, a) = (8.29, 1.96, 1.00, 0.42, 0.12). 
\end{align}

The three kernel models considered in this paper hereafter will be referred to as the `simple' white-noise (or WN) kernel, (Eq.~\ref{eq:kernel_white_noise}), the `standard' kernel (or GP, Eq.~\ref{eq:kernel}) and the `complicated' kernel, (or GPc, Eq.~\ref{eq:kernel_complex}).
The maximum GP log-likelihood (Eq.~\ref{eq:logz}) values (computed over the time range $0M\leq t\leq 200M$) were $4.31 \times 10^6$, $7.34 \times 10^6$ and $ 7.35 \times 10^6$ for the simple, standard, and complicated kernels respectively. 
This suggests the simple kernel is a significantly worse description of the data, and the complicated kernel offers only a small improvement over the standard kernel. 
If Fig.~\ref{fig:gp_shaded_regions} were replotted using this complicated kernel then there would be no noticeable differences in the shaded regions.

The kernels can also be compared using the symmetrized Kullback–Leibler Divergence (KLD) to measure the distances between zero-mean Gaussian distributions with covariance matrices constructed from the three kernels. 
This is plotted in Fig.~\ref{fig:JS_histogram} where it can be seen that that the simple kernel differs significantly from the standard kernel. The standard and complicated kernels are much more similar, as suggested by their similar log-likelihood values. However, the complicated kernel requires an additional three parameters to fit the data which is not penalized when comparing maximum likelihood values. 
We find that the standard kernel is a significant improvement over the simple kernel and is sufficiently flexible to model the waveform uncertainty without overfitting; hereafter, the standard kernel is used for all calculations unless explicitly stated otherwise.

\begin{figure}[t]
    \centering
    \includegraphics[width=0.48\textwidth]{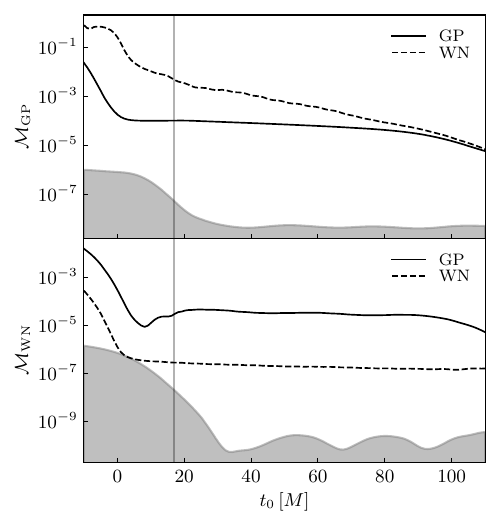}
    \caption{ \label{fig:mismatch} 
        The mismatch of the MAP QNM model compared to the NR data as a function of the ringdown start time for Model 1. The top panel gives the mismatch, calculated with respect to an inverse covariance matrix computed using the standard kernel function ($\mathcal{M}_{\rm GP}$) while the lower panel shows the typical WN mismatch $\mathcal{M}_{\rm WN}$. The vertical black line indicates $t_0=17M$ which is used as the reference start time in later plots. In both panels the gray shaded regions gives an indication of the noise floor of the simulation and show the mismatch between the NR waveforms at the two highest resolution levels.
    }
\end{figure}

%%%
\section{QNM Inference Results} \label{sec:QNMresults}

As an example of the new Bayesian approach to modeling the merger and ringdown, this section shows the results of some QNM fits to an example NR waveform. 
It should be stressed that the purpose of the results in this section is to demonstrate the new method by reproducing known results; readers interested in the new physical insights that can be gained using the new methods are referred to the accompanying paper Ref.~\cite{content_paper}.

As an example, we consider fitting an overtone QNM model to just the $\beta=(2,2)$ mode of the SXS:CCE:0001 simulation. 
The QNM overtone model includes the following QNMs: $(2,2,n,+)$ for $n\in\{0,1,\ldots, 6\}$. The remnant BH mass and spin were also included as free parameters.
This model was chosen because it is well understood, having been previously  investigated by many authors using least-squares methods (see, for example, \cite{Giesler_2019, 2021PhRvD.103h4048F, 2021PhRvD.103j4048D, 2021PhRvD.104l4072F, 2021arXiv211015922M}) and also using QNM filters \cite{2022PhRvD.106h4036M}. However, the methods described here provide additional insight by allowing the study of the full posterior distribution of the QNM model parameters. We call this overtone model `Model 1'. For comparison, we also include results for this overtone model plus the next leading QNM that mixes into the $(2,2)$ spherical mode, $\alpha = (3,2,0,+)$. We call this `Model 2'. 
App.~\ref{app:320qnm} also contains some results with this model.

\begin{figure}[t]
    \centering
    \includegraphics[width=0.48\textwidth]{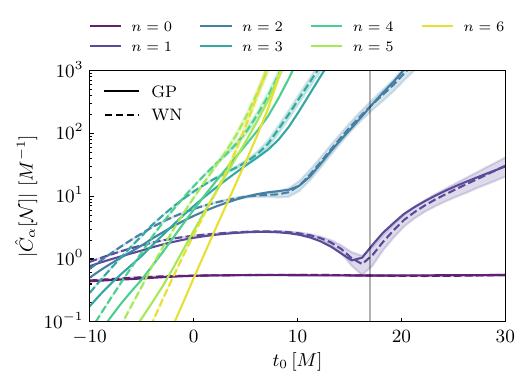}
    \caption{\label{fig:amplitude} 
        The QNM decay-corrected amplitudes $|\hat{C}_\alpha|$ as a function of the ringdown start time for Model 1. 
        Solid (dashed) lines show the results obtained using the GP (WN) covariance and the colors distinguish the QNM overtones.
        The median value is plotted and the shaded regions show the 50\% confidence intervals for the Prior 1 posterior on the WN kernel. The interval is too small to be seen for the GP kernel, so has been omitted. 
        The logarithmic scale used on this plot to show all the overtones gives the misleading impression that the width of the amplitude posteriors is constant at late times; this is related to the behavior of Prior 1 at small amplitudes and is discussed in the main text.
    }
\end{figure}

\begin{figure}[t]
    \centering
    \includegraphics[width=0.48\textwidth]{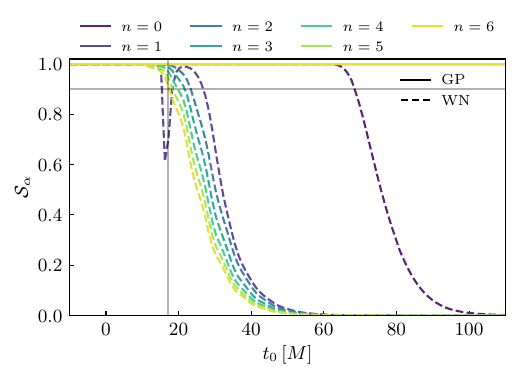}
    \caption{ \label{fig:significance} 
        The significance of each QNM in the model $\mathcal{S}_\alpha$ as a function of the ringdown start time for Model 1.
        Solid (dashed) lines show the results obtained using the GP (WN) covariance and colors are used to distinguish the different QNMs. 
        The vertical line indicates the reference start time $t_0 = 17 M$ and the horizontal line indicates a typical significance threshold of $\mathcal{S}_\alpha = 0.9$, which is approximately equivalent to the 2-dimensional complex amplitude posterior excluding the zero point at two standard deviations.
        }
\end{figure}

For most of the results shown here, the standard (GP) model (Eq.~\ref{eq:kernel}) was used to model the uncertainty in the NR waveform. For comparison, selected results obtained using the simple (WN) model (Eq.~\ref{eq:kernel_white_noise}) are also shown. 
Bayesian fits were performed for ringdown start times in the range $-10 M <t_0<110 M$.
The results of these fits are illustrated in Figs.~\ref{fig:mismatch} to \ref{fig:significance}. On these figures a vertical line indicates a start time of $t_0 = 17M$. This start time was previously found to minimize the angle-averaged mismatch for this simulation; see Table 1 of Ref.~\cite{2025PhRvD.111b4002D}. Subsequent fits are performed at this value of $t_0$. 
Throughout all these plots a consistent color scheme is used to distinguish all the different QNMs in the model and the different line styles are consistently used to distinguish results obtained with the different kernels.

Firstly, the overall quality of the model fit to the data is assessed for Model 1.
This is done by taking the maximum \emph{a posteriori} (MAP) estimate for the QNM parameters and computing the mismatch between a QNM model with these parameters and the NR waveform.
The mismatch between two signals $a$ and $b$
\begin{align} \label{eq:mismatch}
    \mathcal{M}(a,b) = 1-\frac{|\left<a|b\right>|}{\sqrt{\left<a|a\right>\left<b|b\right>}},
\end{align}
can be computed using any suitable inner product $\left<\cdot|\cdot\right>$. 
Here, two inner products are considered: a WN inner product (denoted $\mathcal{M}_{\rm WN}$) and an inner product computed with respect to the inverse covariance matrix defined using the standard GP model described above (denoted $\mathcal{M}_{\rm GP}$).
These mismatches are plotted in Fig.~\ref{fig:mismatch}.

Note, that if the quality of the fit is assessed using the WN mismatch $\mathcal{M}_{\rm WN}$, then the GP results appear to perform worse than the simple results.
This is inevitable because maximizing the QNM likelihood $P(\mathfrak{h}|\boldsymbol{\theta})$ is, by definition, related to minimizing the mismatch.
Similarly, the standard kernel results appear to perform better when judging by the $\mathcal{M}_{\rm GP}$ mismatch. Furthermore, in both cases, the mismatch is determined solely using the MAP value, rather than taking advantage of the full posterior distribution.
Therefore, the waveform mismatch (with either definition) is not the right metric by which to judge the quality of fit. As an alternative, Sec.~\ref{sec:PPC} defines a measure of the quality of the fits based on a Bayesian posterior predictive check (PPC).

Secondly, the QNM mode amplitudes obtained in these fits are examined.
Figure \ref{fig:amplitude} shows the decay-corrected\footnote{
    With our definition of QNM model (see Eq.~\ref{eq:QNMmodel}) the amplitudes $|C_{\alpha}|$ obtained at later start times decay with $t_0$. The decay-corrected amplitude $|\hat{C}_{\alpha}|=|C_{\alpha}\exp(-i\omega_\alpha t_0)|$ corrects for the expected exponential decay by referring the amplitude back to a chosen reference time which is here taken to be $t=0$.
} amplitudes obtained from both the GP and WN fits.
As expected, both the standard and simple versions of the fit obtained essentially identical values for the amplitude of the fundamental mode which is exceptionally stable across the entire range of start times considered.
The amplitudes of the first few overtones show some stability for a smaller range of start times before starting to drift upwards; this behavior has been extensively studied in the literature and is connected to faster decay of the higher overtones which certainly means they are not expected to be detectable at late times.
The high-order overtone amplitudes show no stability with varying start time.

\begin{figure}[t]
    \centering
    \includegraphics[width=0.48\textwidth]{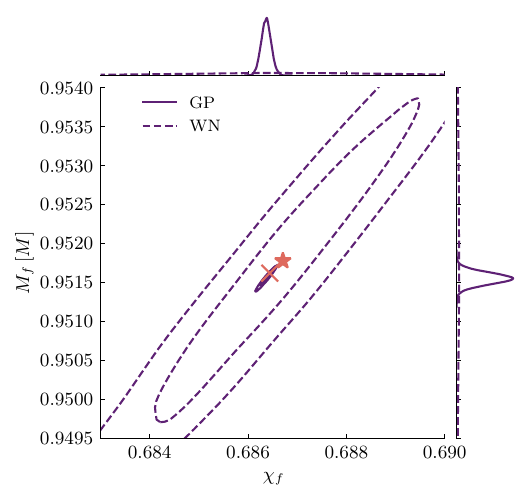}
    \caption{ \label{fig:mass_spin_corner} 
        A joint plot of the two-dimensional marginalized posterior on the remnant mass and spin parameters for QNM fits performed at a start time $t_0 = 17 M$ using Model 2. In the main panel 50\% and 90\% isoprobability contours are shown both for the posterior obtained using the GP kernel (solid lines) and the WN kernel (dashed lines). The cross marks the remnant parameters obtained from the ABD in the NR simulation. The star marks the remnant parameters obtained from the nonlinear least-squares model fit performed prior to the Bayesian inference. Either point could be used as the reference value about which to expand the linear approximation. Throughout this paper we use the ABD value. The posterior obtained using the standard noise model is much more constraining than that obtained using the simple model.
    }
\end{figure}

\begin{figure*}[t]
\begin{minipage}[b]{0.48\linewidth}
\centering
\includegraphics[width=\textwidth]{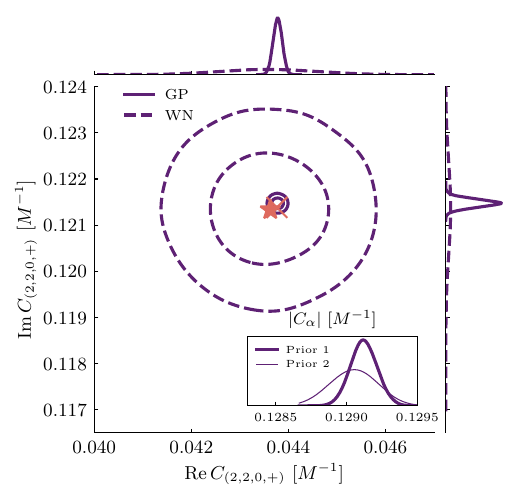}
\end{minipage}
\hspace{0.5cm}
\begin{minipage}[b]{0.48\linewidth}
\centering
\includegraphics[width=\textwidth]{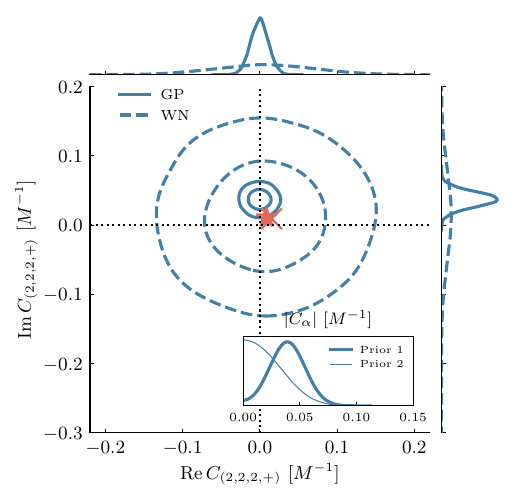}
\end{minipage}
\caption{
    A joint plot of the two-dimensional marginalized posterior on the real and imaginary parts of selected QNM amplitudes $|C_\alpha|$ for QNM fits performed at a start time $t_0 = 17 M$ using Model 2. The left-hand plot shows the fundamental QNM $\alpha=(2,2,0,+)$ while the right-hand panel shows the $n=2$ overtone $\alpha=(2,2,2,+)$. In the main panels 50\% and 90\% isoprobability contours are shown both for the posterior obtained using the GP kernel (solid lines) and the WN kernel (dashed lines). 
    These plots were made using $N_{\rm samples}=10^4$.
    The star markers indicate the point estimates for the amplitudes obtained from the nonlinear least-squares model fit performed prior to the Bayesian inference. The cross marks the linear least-squares fit, with mass and spin fixed to their ABD values.
    The posterior on amplitude of the fundamental QNM (left panel) confidently excludes zero. The fundamental mode has an extremely high significance of $\mathcal{S}_{(2,2,0,+)} \approx 1$ for both the GP and WN kernel posteriors. However, the $n=2$ overtone amplitude posterior does not confidently exclude the zero point indicated by the horizontal and vertical dotted lines on the right-hand plot. 
    The overtone has significance $\mathcal{S}_{(2,2,2,+)}=0.9758$ ($\mathcal{S}_{(2,2,2,+)}=0.0292$) for the GP (WN) posterior. The inset plots show the posteriors on the amplitudes $|C_\alpha|$ of these modes for both the default prior (Prior 1, thick line, flat on $\mathrm{Re}\,C_\alpha$ and $\mathrm{Im}\,C_\alpha$) and reweighted to the alternative prior (Prior 2, thin line, flat on $|C_\alpha|$). The samples used to build the inset KDEs have been symmetrized about zero to enforce $|C_\alpha|\geq 0$. These plots were made using $N_{\rm samples}=1 \times 10^7$ samples to ensure a sufficiently high effective sample number.
    After reweighting, the effective number of samples is $n_{\rm eff}=1.5 \times 10^{3}$.
    The prior reweighting only has a noticeable effect on modes which already have a low significance and for these modes the effect of switching to Prior 2 is to further suppress the mode amplitude.
}
\label{fig:amplitudes_corner}
\end{figure*}

The shaded confidence intervals for the amplitudes in Fig.~\ref{fig:amplitude} reveal a drawback of our chosen prior (Prior 1) which is flat on the real and imaginary parts of the QNM amplitudes. 
This prior assigns no prior probability to a zero value of the amplitude $|C_\alpha|=0$. 
Consequently, looking a Fig.~\ref{fig:amplitude} would appear to suggest that all of the QNMs have amplitude posteriors that confidently exclude zero at all start times. This is definitely not the case. 
One solution to this problem would be to switch to Prior 2 (flat prior on $|C_\alpha|$) and this is explored below (see Fig. ~\ref{fig:amplitudes_corner}).
However, first we show how to define a measure of significance for each QNM using the posteriors already obtained with Prior 1.

Given a two-dimensional marginalized posterior on the real and imaginary part of a QNM amplitude, the significance of that QNM can be defined by how confidently the posterior excludes the point $C_{\alpha}=0$. This significance $\mathcal{S}_\alpha$ is defined in App.~\ref{app:significance}. 
The significance varies in the range $0\leq \mathcal{S}_\alpha\leq 1$ with higher values corresponding to QNMs that are confidently detected in the NR data.

\begin{figure*}[t]
    \centering
    \includegraphics[width=0.96\textwidth]{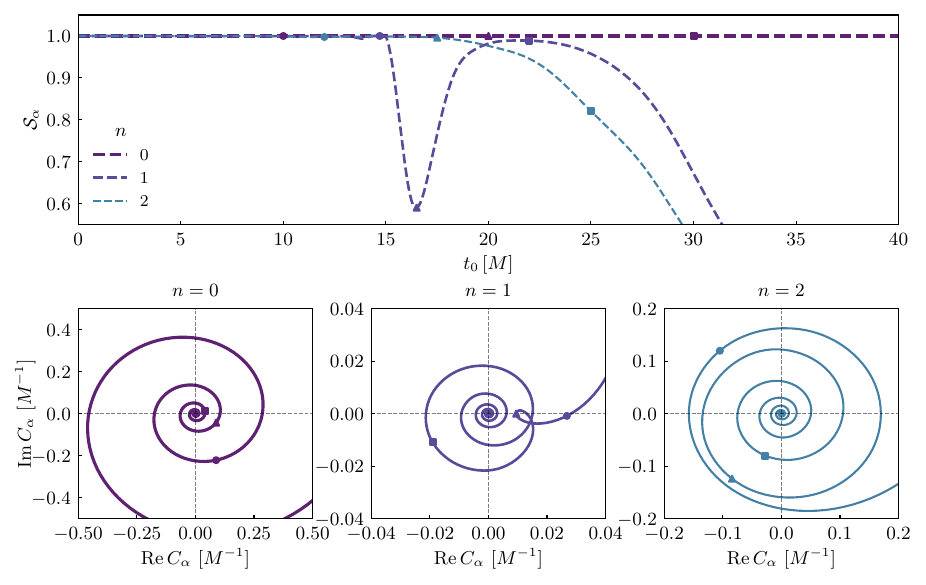}
    \caption{ \label{fig:amplitude_spiral} 
        The significance $\mathcal{S}_\alpha$ curves as functions of start time $t_0$ in Fig.~\ref{fig:significance} show some oscillatory behavior at early times. This is investigated further here for the three QNMs with $n=0$, 1 and 2 using the WN posterior. 
        The significance curves for these modes are reproduced in the top panel.
        The bottom panels show the QNM amplitudes $C_\alpha$ in the complex plane as functions of $t_0$. 
        In both panels, the curves have been smoothed using a cubic spline interpolation.
        The fundamental mode shows the characteristic exponential spiral, corresponding to the predicted decay of the amplitude over the start times considered. The first overtone, which displays the instability, leaves the predicted exponential decay and continues on a new spiral. The behavior here is less regular and the curve self intersects as the amplitude $|C_\alpha|$ increases and decreases. This is the reason for the behavior seen in the top panel and can be compared to the second overtone, which displays a more regular decrease in significance.
        The three markers on each curve indicate specific start times around important features of the curves to enable comparisons between the different panels.
        Animated versions of this figure can be viewed online here: \href{https://github.com/BGP-QNM-FITS/bgp_methods}{\faGithub} \cite{bgp_methods}.
    }
\end{figure*}

Figure \ref{fig:significance} plots the significance $\mathcal{S}_{\alpha}$ for each QNM in the model as a function of the start time for both GP and WN kernels. 
At early times, all the QNM amplitude posteriors are peaked confidently away from zero with high significance, $\mathcal{S}_\alpha \approx 1$.
At late start times the significance of the QNMs drop off in a predictable order, with the higher order overtones decaying first and the fundamental mode surviving longest.
The significance values defined using the standard kernel remain higher than those obtained using the simple kernel for longer; this is related to the fact the GP posteriors are narrower (see, for example, Fig.~\ref{fig:amplitudes_corner}) and much more constraining than the WN posteriors.
At around $t_0 = 16M$, the significance of the second WN overtone fluctuates. This indicates that the amplitude posterior moves closer to zero -- corresponding to a lower significance -- and then further away as $t_0$ changes. The behavior is investigated further in Fig.~\ref{fig:amplitude_spiral} and App.~\ref{app:320qnm}, where it is demonstrated that this may relate to the omission of QNMs from the model, in particular, the absence of the $(3,2,0,+)$ mode. Hereafter, fits are performed using Model 2, which includes this mode. 

Figure \ref{fig:mass_spin_corner} shows the two-dimensional marginalized posterior on the mass and spin at a fixed value of $t_0 = 17M$. 
In both cases, the parameters are well-constrained, which is important because it means the model described in Sec.~\ref{sec:likelihood} is likely to be accurate. This is quantified in App.~\ref{app:linearisedapproximation}.

Figure \ref{fig:amplitudes_corner} shows example posteriors on the complex amplitudes of selected QNMs for $t_0 = 17M$.
For a full set of posterior distributions for every parameter, using the strain, news, and curvature scalar see Appendix~\ref{app:full_param}. 
At this start time the fundamental mode has a high significance, $\mathcal{S_{\alpha}}$, and its posterior is away from zero. The second WN overtone significance has started to decrease at this start time and its amplitude is marginally consistent with zero. For all the QNM amplitudes, the posteriors obtained with the GP kernel are significantly more constraining (i.e.\ narrower) than those obtained with the simple kernel. 

The inset plots in Fig.~\ref{fig:amplitudes_corner} show how the posteriors change when they are reweighted to prior 2. 
This has the effect of decreasing the amplitude of these modes, particularly higher overtones. This decreases the significance of the higher overtones. 

In Fig.~\ref{fig:amplitude_spiral} we revisit the issue of the fluctuating significance values observed in Fig.~\ref{fig:significance}. As the QNM amplitudes $C_\alpha$ decay toward zero, they typically trace out characteristic exponential spirals on the complex plane; the phase advances linearly with start time while the modulus decays exponentially.
However, some deviations from this expected behavior are observed with the amplitudes moving `in and out' over time. These movements are associated with the fluctuating significance values observed above.
The reason for these features is not fully understood but is connected to the fact that Model 1 is incomplete and in particular does not include the $\alpha=(3,2,0,+)$ mode which is known to be important. To demonstrate this, App.~\ref{app:320qnm} shows a version of Fig.~\ref{fig:significance} for Model 2, where $(3,2,0,+)$ has been included. 

%%%
\section{Posterior predictive checking} \label{sec:PPC}

\begin{figure}[t]
    \centering
    \includegraphics[width=0.48\textwidth]{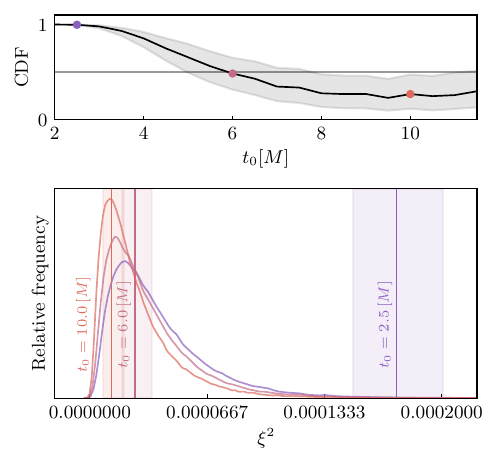}
    \caption{ \label{fig:ppc} 
        \emph{Top panel: } The proportion of the cumulative distribution function (CDF) of the reference generalized chi-squared distribution $\xi^2$ which falls to the left of the samples taken at a given $t_0$. The shaded region gives the 50\% interval. Colored points indicate the selected $t_0$ values plotted in the lower panel, moving left to right with increasing $t_0$. 
        \emph{Bottom panel:} The median values and $50\%$ widths of the residual-squared values obtained from 1000 samples of the posterior fits at values of $t_0$. The reference generalized chi-squared distributions for the selected $t_0$ values are plotted as KDEs in the background. 
    }
\end{figure}

In Sec.~\ref{sec:QNMresults} it was argued that the mismatch (see Eq.~\ref{eq:mismatch} and Fig.~\ref{fig:mismatch}) is not the right metric by which to judge the quality of the QNM model fit to the NR data. 
In a Bayesian context, the appropriate tool for assessing the quality of fit is a posterior predictive check (PPC). 

PPCs involve simulating multiple realisations of the data under the fitted model and comparing to the observed data.
In the present context, this is achieved by drawing QNM model parameters from the posterior $\boldsymbol{\theta}\sim P(\boldsymbol{\theta}|\mathfrak{h})$, computing the QNM model $h^\beta(t;\boldsymbol{\theta})$ in Eq.~\ref{eq:QNM_model_modes}, and forming the residuals with the NR waveform data,
\begin{align}
    R^\beta(t_a;\boldsymbol{\theta}) \equiv \mathfrak{h}^\beta(t_a) - h^\beta(t_a;\boldsymbol{\theta}),
\end{align}
where, for consistency, we continue to use strain notation. However, as before, this procedure is equally valid in the news and Weyl curvature scalar domains.
We then ask: are the model-data residuals consistent with draws from the GP that models the numerical noise?

This question is answered by performing a hypothesis test using the $L^2$ norm of the residuals as a test statistic,
\begin{align}
    \xi^2 \equiv ||R||_{L^2}^2 = \sum_\beta \sum_a  \bigg[ \mathrm{Re}(R^\beta_a)^2 + \mathrm{Im}(R^\beta_a)^2 \bigg] . 
\end{align}
This statistic is used to assess the quality of fit; we use the symbol $\xi$ to avoid confusion with the standard chi-squared distribution and because the symbol $\chi$ is already used to denote the dimensionless BH spin.
The null hypothesis asserts that the residuals are distributed according to the GP used to model the numerical noise,
\begin{align}
    R(t) \sim \mathcal{GP}(0, k) . 
\end{align}
(Hereafter the spherical harmonic $\beta$ index is suppressed in our notation for clarity.)
It is necessary to know the sampling distribution of $\xi^2$ under the null hypothesis.

The $L^2$ norm follows a generalized type of chi-squared distribution. 
In order to show this we use the eigenfunction description of GPs \cite{Rasmussen2006Gaussian}.
Mercer's theorem  allows the kernel to be written as an infinite sum of orthonormal eigenfunctions, 
\begin{align}
    k(t,t') = \sum_i \lambda_i \phi_i(t)\phi_i(t') .
\end{align}
Random functions $f(t)\sim\mathcal{GP}(0, k)$ can also be expressed in this basis using the Karhunen–Lo\`{e}ve expansion,
\begin{align}
    f(t) = \sum_i \sqrt{\lambda_i}Z_i\phi_i(t) , 
\end{align}
where the coefficients are standard normal variates, $Z_i\sim\mathcal{N}(0,1)$. 
Computing the norm gives 
\begin{align} \label{eq:generalised_chi_squared}
    ||f||_{L^2}^2  = \int_{t_0}^T \mathrm{d}t \; \left(\sum_i \sqrt{\lambda_i}Z_j\phi_i(t) \right)^2 = \sum_i \lambda_j Z_i^2 , 
\end{align}
where the second step has used the orthonormality property of the kernel eigenfunctions.
This shows that the squared norm follows a type of generalized chi-squared distribution with the eigenvalues $\lambda_i$ playing the role of weights. 
(Recall, for standard chi-squared random variable with $\nu$ degrees of freedom the sum in Eq.~\ref{eq:generalised_chi_squared} would contain only $\nu$ terms all equally weighted with $\lambda_i=1$.)

The eigenvalues $\lambda_i$ can be computed numerically from the Gram matrix $\boldsymbol{K}_{ab}=k(t_a,t_b)$ on the discretely sampled on the discrete grid of times $t_a$.

Figure \ref{fig:ppc} shows the results of a PPC for the $i=\,$0001 simulation modeled using the GP kernel and Model 2, at a range of start times.
The model-data residuals were calculated for 1000 sets of QNM parameters drawn from the posterior distribution.
For each set of residuals the statistic $\xi^2$ was calculated and the distribution of these values is shown in the vertical shaded bands in the bottom panel.
These observed values are then compared to the generalized chi-squared sampling distribution in Eq.~\ref{eq:generalised_chi_squared}; this is shown using KDEs in the bottom panel. 
Note that this sampling distribution depends somewhat on the choice of $t_0$ through its dependence on the eigenvalues of $\boldsymbol{K}$.
A $p$-value-type quantity can then be defined by computing the cumulative distribution function (CDF) of the sampling distribution at the observed value of the statistic. 
Large values of the CDF indicate that the model lies in the right-hand tail of the distribution and provides a poor fit to the data.
Conversely, small values of the CDF indicate that the model lies in the left-hand tail of the distribution and is overfitting the data.
CDF values $\sim 1/2$ indicate a good fit.
The results in Fig.~\ref{fig:ppc} show that the combined QNM and GP model underfit at early times ($t_0\lesssim 5M$) and begin to overfit at late times ($t_0\gtrsim 8M$). This can be compared to Fig.~\ref{fig:ppc_WN} in App.~\ref{app:wn_ppc} which shows the same PPC using a WN model.

%%%
\section{Discussion and Concluding Remarks} \label{sec:discussion}

In this work, a Bayesian framework for fitting QNMs to CCE waveforms has been introduced, incorporating a GP model to account for numerical uncertainty in the NR data. For a simple overtone model, the method reproduces many of the familiar results reported in earlier works.

The method can be applied to the strain, news, or the curvature scalar. When applied to the latest CCE waveforms, we show that the best results are obtained from the news or curvature scalar, owing to their decay to zero at late times. 

The method is an improvement over least-squares fitting because it includes a more reliable numerical uncertainty estimate, and provides posterior distributions rather than point estimates. Furthermore the physically-motivated noise model described in the paper leads to significantly tighter posteriors than those obtained under as WN treatment, demonstrating the value of constructing a well-motivated model and training this on a catalog of waveforms. 

We have shown that our particular choice of kernel is well-suited to capturing numerical uncertainty in the CCE waveforms. While already flexible, the kernel can be extended further. For example, it could also describe the inspiral and merger portions of the NR signal, or be applied to extrapolated simulations. 

The kernel could also be applied to larger training sets. The cost of training scales linearly with the number of simulations included in the training. While the methods here could be further optimized, the training is a one-off cost, so it would be relatively easy to scale to significantly more simulations from, for example, the SXS catalog.

A particular strength of the Bayesian fitting approach is in its computational efficiency. The linearized structure of the model means that, in addition to the amplitudes, the mass and spin posteriors can be obtained extremely rapidly without reliance on expensive MCMC or other stochastic sampling methods. The method therefore scales efficiently to models with arbitrary numbers of QNMs, with a computational cost comparable to least-squares fitting. Flexibility in prior choice is also achieved, since alternative priors can be accommodated by rapid sample reweighting. 

This Bayesian approach will also reduce ambiguity surrounding (sometimes contentious) claims of subdominant mode detections. We have outlined a new quantity called the significance, which gives the consistency of a mode amplitude with zero. The presence of specific QNMs in the ringdown can now be determined using this statistical measure. 

Finally, a Bayesian PPC has been described, allowing assessment of whether a given QNM model under- or over-fits the data. Here, the PPC suggests that a many overtone model (at least up to $n=6$) provides a good fit at early start times but quickly begins to overfit ($t_0 \geq 6M$) the data. This diagnostic offers a natural means of probing the QNM content of the ringdown and building more realistic models, which is explored further in \cite{content_paper}.

%%%%%%%%%%%%%%%%%%%%%%%%
%%% Acknowledgements %%%
%%%%%%%%%%%%%%%%%%%%%%%%

\section*{Data and Code Availability}

The code developed to perform the Bayesian fits to NR waveforms has been made publicly available at this repository \href{https://github.com/BGP-QNM-FITS/bgp_qnm_fits}{\faGithub}~\cite{bgp_qnm_fits}. 
All the results and plotting scripts have additionally been made publicly available at this repository \href{https://github.com/BGP-QNM-FITS/bgp_methods}{\faGithub}~\cite{bgp_methods}. 
We are grateful that the NR data used in this study have already been make publicly available \cite{SXS_CCE_catalog} by the SXS collaboration and this data was processed using the methods described in the main paper.

\begin{acknowledgments}
    We are grateful to Eliot Finch, Maximiliano Isi, Keefe Mitman, Erin Coleman, Teagan Clarke, Nicole Khusid, and Harrison Siegel for helpful discussion about this work.
    The authors would also like to thank the anonymous reviewers for their constructive reports which helped improve this manuscript. R. D. acknowledges support from the Science
    and Technology Facilities Council (STFC) through a PhD studentship (Grant reference: ST/Y509139/1).
    
    In addition to the software mentioned in this paper, we also made use of the following packages throughout: \texttt{Numpy} \cite{harris2020array}, \texttt{SciPy} \cite{2020SciPy-NMeth}, \texttt{Matplotlib} \cite{Hunter:2007}, \texttt{JAX} \cite{jax2018github}, \texttt{Seaborn} \cite{Waskom2021}, \texttt{Pandas} \cite{reback2020pandas}, \texttt{qnmfits} \cite{qnmfits_software}, and \texttt{sxs} \cite{sxs_software}.
\end{acknowledgments}

\clearpage

%%%%%%%%%%%%%%%%%%%%
%%% Bibliography %%%
%%%%%%%%%%%%%%%%%%%%

\bibliographystyle{apsrev4-1}
\bibliography{references}

%%%%%%%%%%%%%%%%%%
%%% Appendices %%%
%%%%%%%%%%%%%%%%%%

\appendix

%%%
\section{Accuracy of the linearized approximation in $\chi_f$ and $M_f$} \label{app:linearisedapproximation}

\begin{figure}[t]
    \centering
    \includegraphics[width=0.48\textwidth]{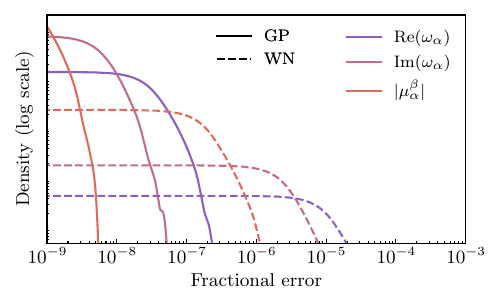}
    \caption{ \label{fig:linear_approx_accuracy}
        KDEs for the fractional errors in selected quantities involved in the linearized approximation to the QNM model. 
        The three quantities considered are the real and imaginary parts  of $\omega_{\alpha}$ with $\alpha=(2,2,0,+)$ and the modulus $|\mu_\alpha^\beta|$ of the primary mode mixing coefficient with $\beta=(2,2)$.
        The solid (dashed) lines show the results for the posteriors obtained using the GP and WN kernels respectively.
    }
\end{figure}

The linearization in the remnant mass and spin parameters, $M_f$ and $\chi_f$, used in the QNM model (see Eq.~\ref{eq:simple_model_linearised}) necessarily introduces some error.
This error is expected to be small because the remnant parameters are typically extremely well constrained in a QNM fit of an NR waveform; see, for example, Fig.~\ref{fig:mass_spin_corner}.
The accuracy of this linear approximation is quantified by drawing samples $M_f,\chi_f$ from the two-dimensional posterior on the remnant parameters.
For each of these posteriors samples any of the QNM frequencies $\omega_\alpha$ or mode mixing coefficients $\mu_\alpha^\beta$ can be calculated in two ways: (i) using the full nonlinear expressions for the Kerr metric (calculated using the \texttt{qnm} package \cite{Stein:2019mop}), or (ii) using the linearized expressions expanded about the reference parameters $\boldsymbol{\theta}_*$ (see, e.g., Eq.~\ref{eq:simple_model_linearised}).
The fractional error between these two quantities can then be calculated and used to quantify the accuracy of the linear approximation.
This was done for three quantities: the real and imaginary parts of the fundamental QNM frequency $\omega_{(2,2,0,+)}$ and the modulus of the most significant mode mixing coefficient $\mu_{(2,2,0,+)}^{(2,2)}$.
The fractional error for a quantity $Y$ is defined as 
\begin{align}
    \frac{|Y_i(\textrm{linear approx.})-Y_i(\textrm{nonlinear})|}{Y_i(\textrm{linear approx.})} ,
\end{align}
and is computed for $10^4$ posterior samples (indexed by $i$) drawn from the distributions shown in Fig.~\ref{fig:mass_spin_corner}.
This was done using both the GP and WN posterior distributions; the WN posterior covers a larger region of the mass-spin plane and therefore the typical errors in the linear approximation are larger in this case.
This can be seen in the results in Fig.~\ref{fig:linear_approx_accuracy} where kernel density estimators (KDEs) are used to show the distributions of these fractional errors. The figure shows that when expanding about the ABD remnant parameters the fractional error in the linear approximation to the QNM model is less than around 1 part in $\sim 10^7$ across the typical set of the GP posterior. The approximation can be further improved by expanding about the nonlinear mass and spin parameters, rather than the ABD values, however such an improvement has a negligible effect on the overall fits.

%%%
\section{QNM significance} \label{app:significance}

A significant drawback of the least-squares approach to fitting QNMs is that it only provides a point estimate of the mode amplitudes and it is not possible to tell from this alone whether or not a particular QNM belongs in the model.
There are several potential reasons why a QNM might not belong in the model.
As QNMs decay, their amplitudes at late start times become comparable to numerical noise and they can no longer be confidently detected.
Conversely, some modes (e.g.\ high-order overtones) have large amplitudes at early start times that destructively interfere with other modes, leaving their physical significance unclear.
The problem of determining the QNM content has been addressed in a variety of \emph{ad hoc} ways. 
For example, fits can be performed for a range of start times to check that the recovered amplitudes are stable \cite{Giesler_2019, zhu2024imprintschangingmassspin}, the real and imaginary parts of a particular QNM frequency can be allowed to vary freely and be compared with predicted QNM frequencies \cite{Giesler_2019, 2023PhRvL.130h1401C, Mitman_2023}, or loud QNMs can be filtered out to identify subdominant modes \cite{2022PhRvD.106h4036M, lu2025statisticalidentificationringdownmodes}.

A key advantage of the Bayesian approach is that it gives a posterior on the QNM amplitudes, not just a point estimate.
The posterior allows us to assess if zero amplitude (i.e.\ no QNM) is consistent with the data. 
Furthermore, this can be done from a single fit, without the need to vary the start time to check for amplitude stability or introduce additional parameters as in a free frequency fit.

For a particular QNM, indexed by $\alpha$, we introduce the following measure for its significance ($0\leq \mathcal{S}_{\alpha}\leq 1$);
\begin{align} \label{eq:QNMsignificance}
    \mathcal{S}_{\alpha} = 1 - \exp\left(-\dfrac{1}{2}|b_{\alpha}|^2\right),
\end{align}
where $\mu_\alpha$ and $\Gamma^{-1}_\alpha$ are the mean vector and covariance matrix for the 2-dimensional marginalized posterior on the real and imaginary parts of the QNM amplitude, $L_\alpha$ is the Cholesky decomposition of $\Gamma^{-1}_\alpha$ (i.e.\ $\Gamma^{-1}_{\alpha}=L_\alpha\cdot L_\alpha^{\rm T}$ with $L_\alpha$ lower triangular), and $b_\alpha=L_\alpha^{-1}\cdot\mu_\alpha$.
High $\mathcal{S}_{\alpha}$ indicate a preference for the inclusion of the $\alpha$ QNM; conversely, low $\mathcal{S}_{\alpha}$ indicates a preference for omitting the mode.
The definition of significance in Eq.~\ref{eq:QNMsignificance} can be motivated in either a Bayesian or a frequentist way.

\begin{figure}[t]
    \centering
    \includegraphics[width=0.35\textwidth]{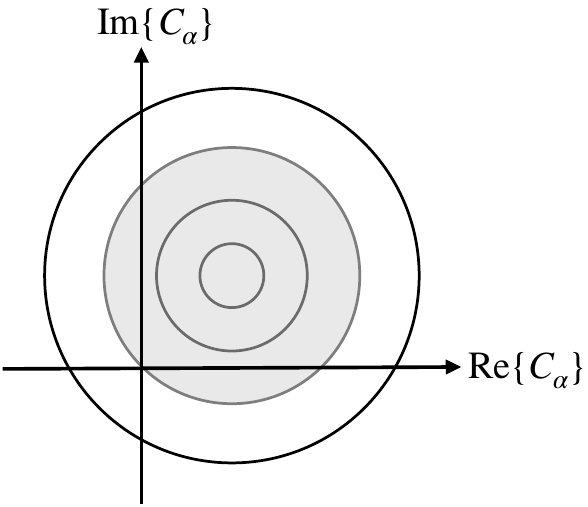}
    \caption{ \label{fig:significance_diagram}
        Diagram illustrating the definition of the significance $\mathcal{S}_\alpha$ in Eq.~\ref{eq:QNMsignificance}. 
        Lines are iso-probability contours of the 2D marginalized posterior on the real and imaginary parts of the QNM amplitude. 
        The integral is over the shaded region.
    }
\end{figure}

In a Bayesian spirit, compute the posterior odds ratio $\mathcal{O}_\alpha$ for a pair of models, with and without QNM $\alpha$, but otherwise identical.
A uniform prior is used on the real and imaginary parts of the QNM amplitude in the range $(-A_{\rm max}, A_{\rm max})$.
As the models are nested, the evidence ratio can be computed using the Savage-Dickey density ratio.
Assuming equal prior odds for the two models, the posterior odds ratio in the limit of large $A_{\rm max}$ is given by 
\begin{align} \label{eq:dropout_Bayes}
    \mathcal{O}_\alpha = \frac{\exp(-\frac{1}{2}|b_\alpha|^2)}{8\pi |L_\alpha| A_{\rm max}^2} ,
\end{align}
As usual in Bayesian inference, the result depends on our prior; in this case, on $A_{\rm max}$.
However, the power-law dependence of the result on $A_{\rm max}$ in the denominator is usually dominated by the exponential dependence on the the data through $b_\alpha$. 
Keeping only the exponential terms we arrive at the approximate result $\mathcal{O}_\alpha\propto 1-\mathcal{S}_\alpha$.

In a frequentist spirit, the significance of the QNM $\alpha$ is defined using the number of standard deviations away from zero the (two-dimensional) posterior on the (complex) mode amplitude is peaked. 
This significance is defined as the fraction of the posterior mass associated with higher posterior density that the point of zero amplitude; i.e.\ the integral of the posterior inside the iso-probability contour that passes through the zero amplitude point $C_\alpha=0$ (see Fig.~\ref{fig:significance_diagram}),
\begin{align}
    \mathcal{S}_\alpha = \int_{\{C_\alpha|P(C_\alpha|d)>P(0|d)\}} \; \mathrm{d}(\mathrm{Re}\,C_\alpha)\; \mathrm{d}(\mathrm{Im}\,C_\alpha) \; P(C_\alpha|d) ,
\end{align}
where $P(C_\alpha|d)$ is the 2D marginalized posterior on the complex amplitude $C_\alpha$.
Integrating the two-dimensional Gaussian posterior on the QNM amplitude with mean $\mu_\alpha$ and covariance $\Gamma_\alpha^{-1}$ gives the result in Eq.~\ref{eq:QNMsignificance}.

For loud QNMs with $\mathcal{S}_\alpha \rightarrow 1$, Eq.~\ref{eq:QNMsignificance} becomes numerically unstable. 
In this limit it is convenient to use
\begin{align}
    \log\mathcal{S}_{\alpha} \approx - \exp\left(-\dfrac{1}{2}|b_{\alpha}|^2\right)\;.
\end{align}

%%%
\section{NR waveform preprocessing} \label{app:data_processing}

The results in this paper use the 13 CCE waveforms provided in the SXS catalog. 
For these simulations, preprocessing described in the main text is carried out. For each simulation, the highest resolution level (L5), extracted at the second smallest world-tube radius, is used as the preferred waveform. 
The same simulation, extracted at the second highest level (L4) is used as the second `best' simulation for computing the residual. 
This was the simulation whose mismatch with the preferred waveform was minimum. 
Therefore it most closely resembles the preferred waveform up to differences due to numerical uncertainty.

The subsequent processing is also applicable to the extrapolated simulations in the main SXS catalog. 
To compute the residual, we first perform a time and a phase shift to align the two waveforms. 
The data is up-sampled using a cubic spline implemented in \texttt{scipy.interpolate.make\_interp\_spline}. 
The choice of interpolation timestep is taken from \cite{2025PhRvD.111b4002D}, chosen so that the change in mismatch due to rolling the wavefunction by a timestep has converged to a negligible value. 
The shift can then be computed by maximizing the overlap between the two waveforms. 
This was done efficiently using fast Fourier transform techniques.

To compute the residual, the time-shift corrected waveforms are re-interpolated onto the same (coarser) time grid and subtracted from each other; 
\begin{align}
    \mathfrak{r}^\beta_i = \mathfrak{h}^\beta_i(\mathrm{L5}) - \mathfrak{h}^\beta_i(\mathrm{L4}).
\end{align}
These residuals are used to train the GP model for the waveform uncertainties (Sec.~\ref{sec:GPcov}). 

\begin{figure}[t]
    \centering
    \includegraphics[width=0.5\textwidth]{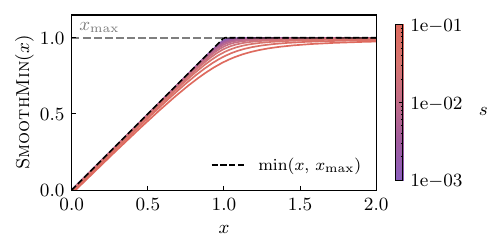}
    \caption{ \label{fig:smoothmin}
    The $\text{\textsc{SmoothMin}}(x, x_{\rm max}; s)$ function (Eq.~\ref{eq:sclip}) with $x_{\rm max}\!=\!1$ and $10^{-3} \leq s\!\leq\! 10 ^{-1}$. As $s\!\rightarrow\! 0$ it approaches the black dashed line with discontinuous first derivative.
         }
\end{figure}

%%%
\section{The SmoothMin function} \label{app:smoothmin}

The \textsc{SmoothMin} function used in the definition of the GP kernel functions in Eqs.~\ref{eq:kernel} and \ref{eq:kernel_complex} is defined as 
\begin{align} \label{eq:sclip}
    \text{\textsc{Smooth}}&\text{\textsc{Min}}(x,x_{\rm max};s) = \\ \nonumber &\frac{1}{2}\Bigg[x + x_{\rm max}\Bigg(1 - \sqrt{\Big(\frac{x}{x_{\rm max}} -1\Big)^2 + s}\Bigg) \Bigg].
\end{align}
This is a smooth version of the $C^0$ function $\mathrm{min}(x; x_{\rm max})$ with a parameter $s>0$ that controls the smoothness.
With this definition, in particular with $x_{\rm max}$ factored outside the parentheses containing the square root, the effect of the smoothing parameter is independent of the scale $x_{\rm max}$.
The function is plotted in Fig.~\ref{fig:smoothmin}.

%%%
\section{Regularization of the GP} \label{app:reg_gp}

\begin{figure*}[t]
    \centering
    \includegraphics[width=0.99\textwidth]{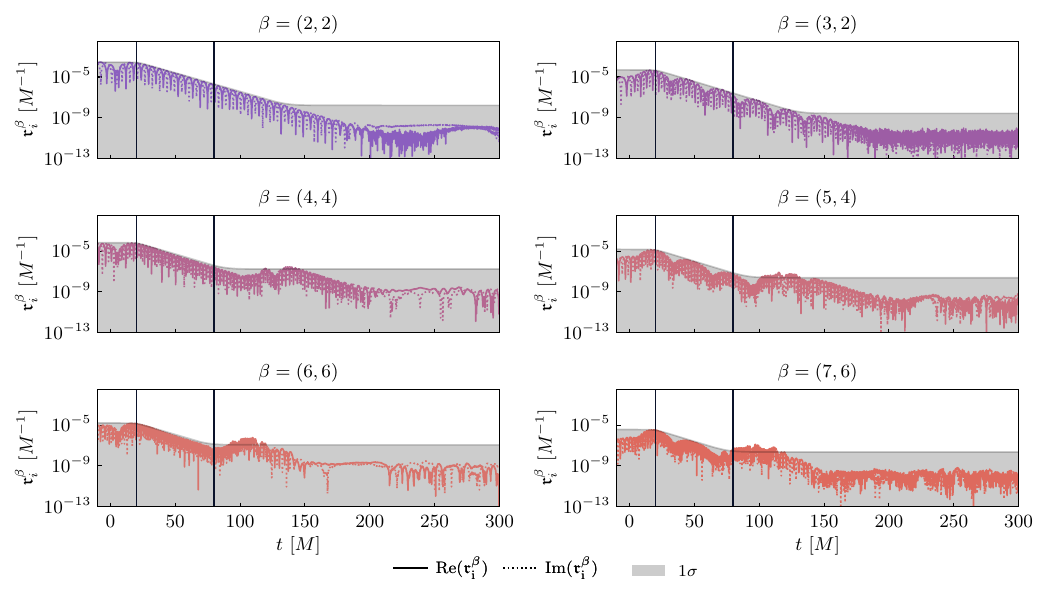}
    \caption{ \label{fig:gp_shaded_regions_logarithmic}
        A version of Fig.~\ref{fig:gp_shaded_regions} from the main text plotted using a logarithmic $y$-axis scale, with a plot range extended to later times and with more spherical modes included. This highlights the role played by the regularizing $\epsilon$ parameter. The bump around $100 - 130 \,\, M$ that is visible in the $\ell \geq 4$ spherical modes is likely related to a re-gridding that occurs in the adaptive mesh refinement (AMR) \cite{keefe_priv}.
    }
\end{figure*}

Figure \ref{fig:gp_shaded_regions_logarithmic} shows the role of the jitter term on controlling the late-time form of the GP kernel. In particular, the figure demonstrates how the value of $\epsilon$ sets the height on the late-time uncertainty. A higher value of $\epsilon$ would increase the height of the flat region (which for $\epsilon = 10^{2}$ starts at $t \sim 120 M$ in the $(2,2)$ mode) this would consequently reduce the time at which this part of the model dominates. On the other hand, a smaller value of $\epsilon$ pushes the end of the exponential decay segment to later times. However numerical instabilities start to arise as the value becomes too small. A value of $\epsilon = 10^{2}$ was chosen to balance conservatively modeling the late-time uncertainty and ensuring numerical stability, while accurately describing the residual in this regime.  

%%%
\section{Sampling frequency checks} \label{app:sampling_frequency}

\begin{figure}[t]
    \centering
    \includegraphics[width=0.48\textwidth]{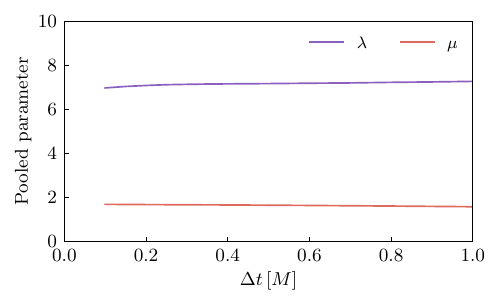}
    \caption{ \label{fig:hyperparameters_dt} 
        The pooled parameter values as a function of timestep $\Delta t$. These values demonstrate good stability. 
    }
\end{figure}

To ensure that the sampling densities used in training the GP model, and used to perform the fits, are not sensitive to the timestep $\Delta t$, we perform two stability tests. The first test shown in Fig. \ref{fig:hyperparameters_dt} gives the values of the pooled parameters across a range of sampling time steps, demonstrating good stability for $\Delta t < 1$. The small amount of drift in these values we attribute to small numerical instabilities as the matrices become significantly larger.

\begin{figure}[ht]
    \centering
    \includegraphics[width=0.48\textwidth]{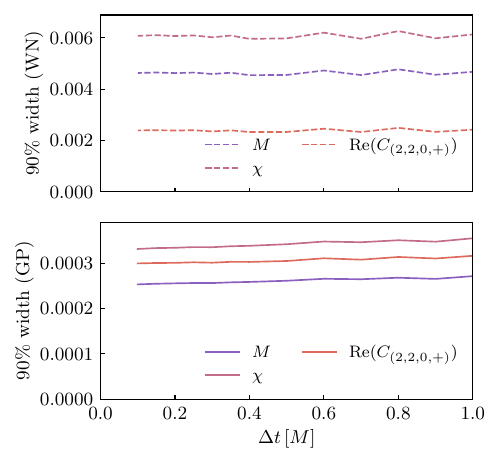}
    \caption{ \label{fig:parameters_dt} 
        Three model parameters as a function of timestep (with fixed kernel parameters). The WN parameters are stable across a range of $\Delta t$. The GP parameters are also relatively stable, with some deviation attributed to the necessary addition of jitter.
    }
\end{figure}

In Fig.~\ref{fig:parameters_dt} we fix the pooled parameters to their values at $\Delta t = 0.1$, and perform fits on data sampled at a range of timesteps. Both the WN and GP kernels display relatively good stability across a range of values of $\Delta t$. In the GP case, there is some drift, which can be primarily attributed to the jitter term. As $\Delta t$ decreases, the jitter has more of an impact on the eigenvalues of the covariance matrix. The mixing of a WN-like kernel with the GP kernel means the inner product used to compute the fisher matrix does not perfectly converge, resulting in a drift. As the jitter is necessary for numerical stability, this is unavoidable. Smaller values of $\epsilon$ will reduce this effect, but result in greater numerical instability. 

While both tests vary the sampling frequency of the data, note that the timestep $\Delta t$ used in training need not be the same as that used for fitting. 

\begin{figure}[t]
    \centering
    \includegraphics[width=0.48\textwidth]{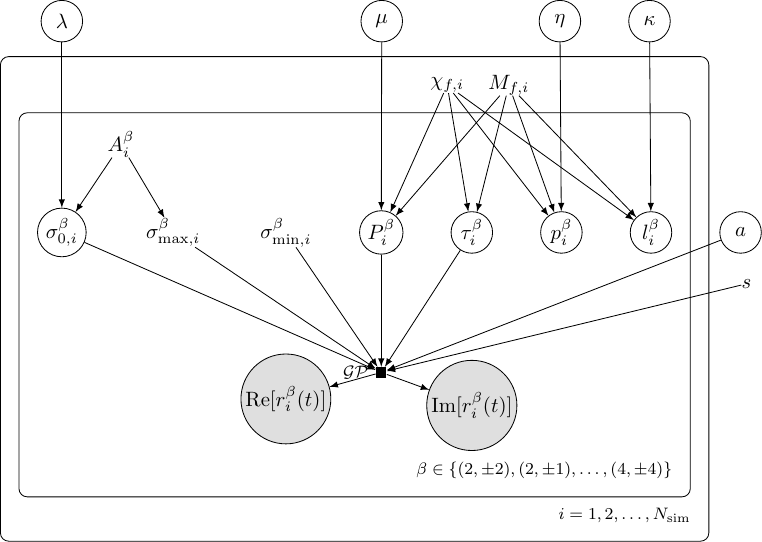}
    \caption{ \label{fig:pgmgpc}
        PGM of the complicated GP model in Eq.~\ref{eq:kernel_complex} for the waveform uncertainty.
        This should be compared with the PGM for the GP kernel model in Eq.~\ref{eq:kernel} shown in Fig.~\ref{fig:pgm} of the main paper.
    }
\end{figure}

%%%
\section{The complicated (GPc) kernel} \label{app:GPC}

The main GP model (Eq.~\ref{eq:kernel}) for the waveform uncertainty was illustrated using the PGM in Fig.~\ref{fig:pgm} in the main body of the paper. 
Here the PGM for the complicated (GPc) model (Eq.~\ref{eq:kernel_complex}) is shown in Fig.~\ref{fig:pgmgpc}.

%%%
\section{The stationary kernel} \label{app:wendland}

The stationary kernel introduced in the main text in Sec.~\ref{sec:GPcov} is defined to be the one-dimensional, $q=3$ Wendland covariance function \cite{Rasmussen2006Gaussian};
\begin{align}
    k_{\rm Stationary}(t, t'; P_{i}^{\beta}) = k_{\rm Wendland}\left(\frac{|t-t'|}{P_{i}^{\beta}}\right) ,
\end{align}
where
\begin{align}
    k_{\rm Wendland}(\tau) = 
        &\frac{1}{15}(1 - \tau)_{+}^{j+3} \times \\
        \big(&(j^{3} + 9j^{2} + 23j + 15) \tau ^{3} + \nonumber\\
        &(6j^{2} + 36j + 45) \tau ^{2} + \nonumber\\
        &(15j + 45) \tau + \nonumber\\
        &15\big)  , \nonumber
\end{align}
and where $j = q + 1$ and $(x)_+\equiv\mathrm{max}(0,x)$.

%%%
\section{Including the (3,2,0,+) QNM} \label{app:320qnm}

To highlight the limitations of the overtone model (Model 1) discussed in this paper, we reproduce Fig.~\ref{fig:significance} from the main text, but with the loudest $(\ell, m) \neq (2, 2)$ mode included. 
This is the fundamental prograde QNM with $(\ell, m)=(3,2)$ which appears in the $\beta=(2,2)$ spherical harmonic strain as a result of mode mixing. This model is referred to as Model 2 in the main text. 

\begin{figure}[t]
    \centering
    \includegraphics[width=0.48\textwidth]{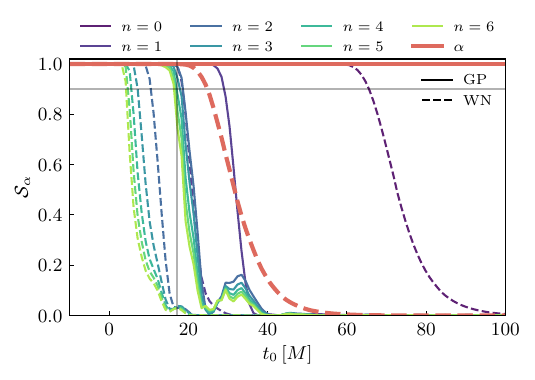}
    \caption{ \label{fig:significance_320} 
        A version of Fig.~\ref{fig:significance} for the QNM model including the $\alpha=(3,2,0,+)$ mode (i.e.\ Model 2).
        The significance of each QNM in the model $\mathcal{S}_\alpha$ as a function of the ringdown start time.
        Solid (dashed) lines show the results obtained using the GP (WN) covariance and different colors are used to distinguish the different QNMs.
    }
\end{figure}

Figure \ref{fig:significance_320} displays the same characteristics as in Model 1, with overtones dropping in significance more rapidly with increasing overtone number. 
However, the spike in the significance at early start times has disappeared. 
We also note that the $\alpha=(3,2,0,+)$ QNM is the second longest lived mode in the fit, and dominates over the overtones for a substantial range of start times. However, it is unlikely that the $(3,2,0,1)$ mode is as significant in $(2,2)$ as this figure alone would suggest. 
In practice, a multimode fit including the $(3,2)$ mode would be required to give a more accurate representation of the mode's contribution to the ringdown. 

%%%
\section{Comparing residuals from the strain, news and curvature scalar} \label{app:residual_comparison}

\begin{figure}[t]
    \centering
    \includegraphics[width=0.48\textwidth]{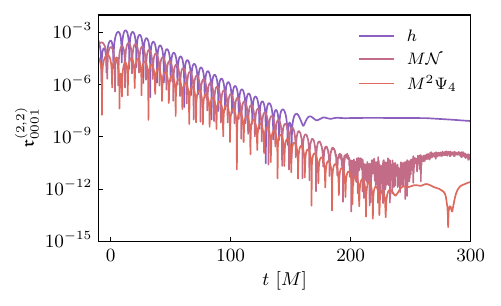}
    \caption{ \label{fig:residual_comparison}
    A comparison of the residuals for the $\beta = (2,2)$ mode in the $i = 0001$ simulation for the strain, news, and curvature scalar. The late-time residual of the strain is clearly much larger than the news and curvature scalar, resulting from imperfections in the superrest frame mapping.
    }
\end{figure}

In this paper, we chose to work with the news rather than the strain, which is typically what is used in ringdown studies. This is due to imperfections that appear after the superrest frame transformation of the strain and result in a much larger late-time residual. These do not appear in the news, which always decays to zero regardless of the frame. This can be seen in Fig.~\ref{fig:residual_comparison}. The imperfection inflates the value of $\sigma_{\mathrm{min},i}^{\beta}$ compared to the news and curvature scalar which results in a greater estimate of the uncertainty and hence a much less well-constrained posterior. 

%%%

\section{White noise posterior predictive check} \label{app:wn_ppc}

\begin{figure}[t]
    \centering
    \includegraphics[width=0.48\textwidth]{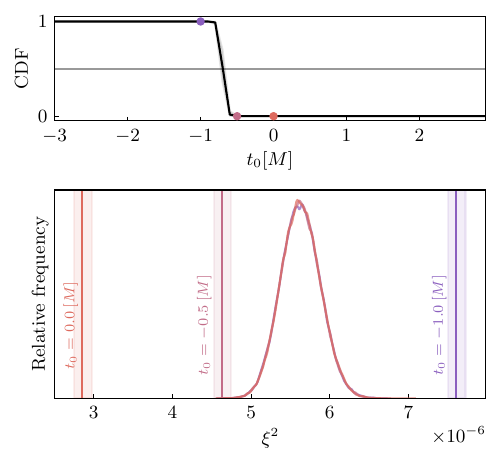}
    \caption{ \label{fig:ppc_WN}
    An example of the WN PPC applied to Model 2. 
    \emph{Top panel: } The proportion of the cumulative distribution function (CDF) of the reference generalized chi-squared distribution $\xi^2$ which falls to the left of the samples taken at a given $t_0$. Colored points indicate the selected $t_0$ values plotted in the lower panel, moving left to right with increasing $t_0$. The 50\% interval is too narrow to be clearly seen. 
    \emph{Bottom panel:} The median values and $50\%$ widths of the residual-squared values obtained from 1000 samples of the posterior fits at values of $t_0$. The reference generalized chi-squared distributions (which are almost identical across the $t_0$ values chosen here) are plotted as KDEs in the background. 
    }
\end{figure}

Fig.~\ref{fig:ppc_WN} shows the WN PPC for Model 2. This figure should be compared with Fig.~\ref{fig:ppc}, which shows the PPC determined when using the GP. The WN model demonstrates clear overfitting at much earlier start times than for the GP, indicating that it does not perform as well at modeling the NR uncertainty.  

%%%
\section{Comparing posteriors from fitting the strain, news and curvature scalar} \label{app:full_param}

\begin{figure*}[t]
    \centering
    \includegraphics[width=0.8\paperwidth]{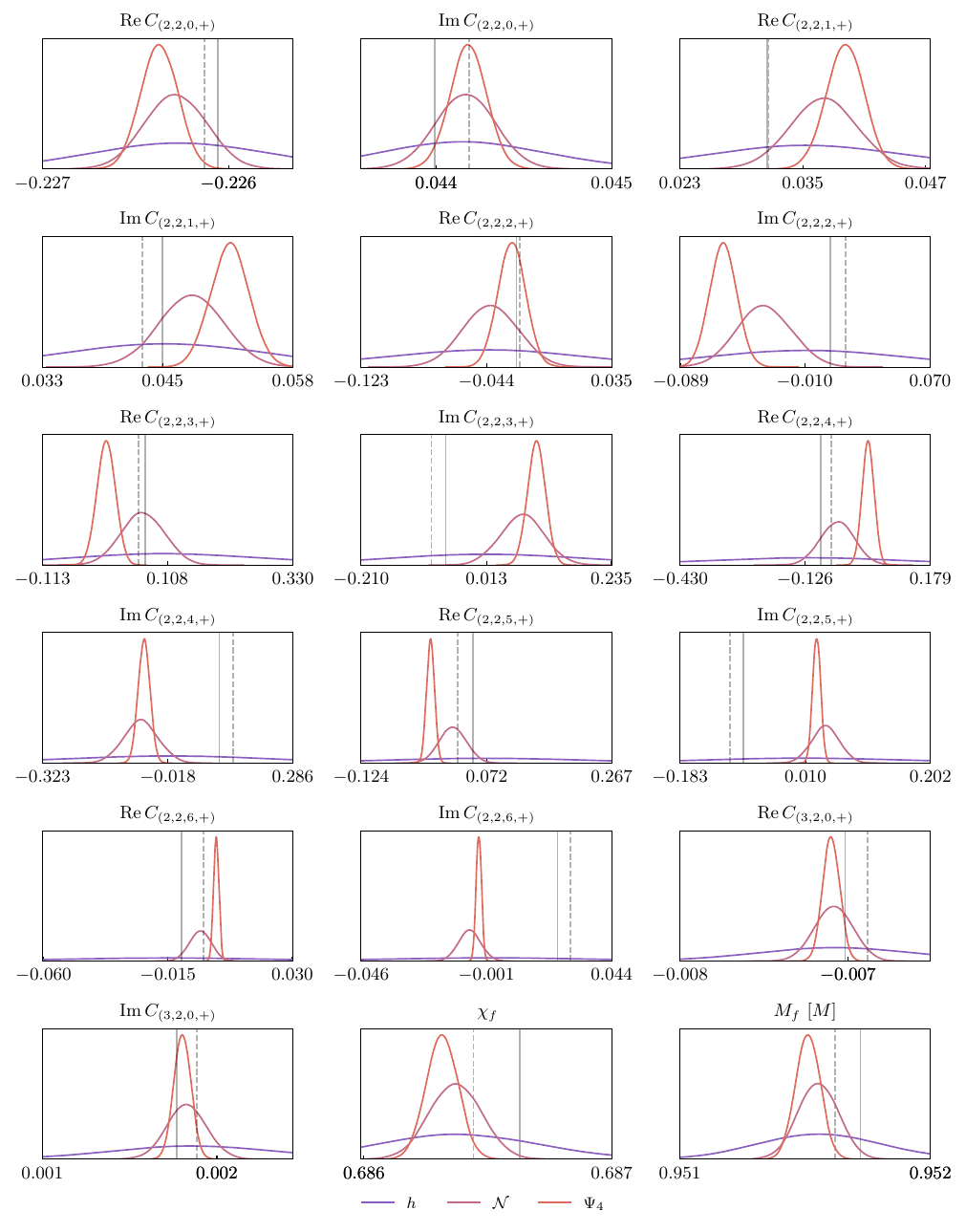}
    \caption{ \label{fig:strain_news_psi4} 
        The posterior distributions for every model parameter in Model 2, on the strain, news, and curvature scalar. The amplitudes have been converted into their strain-domain values. The bold vertical line show results obtained from a non-linear least-squares fit to the strain. The dashed lines show the reference ABD values of the mass and spin and, for the amplitudes, the linear least-squares fits to the strain. 
    }
\end{figure*}

The methods discussed in this paper are equally applicable to the strain, news, and curvature scalar. This is demonstrated in Fig.~\ref{fig:strain_news_psi4} where the full posterior distributions for each parameter in Model 2 at $t_0 = 17M$ are given for each of these quantities. 

In order to compare the fits, the news and curvature scalar amplitudes are converted to the strain using Eq.~\ref{eq:domains}. The linear and non-linear least squares fits to the strain are also shown as vertical lines. In principle, these fits should yield consistent results and this is checked here. 

In general, the MAP values are in broad agreement across the three quantities. However, the posterior widths are not consistent. This can be attributed to differences in the GP noise model and, in particular, the late-time jitter. 
For the strain, which has a larger late-time uncertainty, this jitter is significantly higher, which leads to the wider posterior (see App.~\ref{app:residual_comparison}). The differences in the news and curvature scalar can also be attributed to differences in this part of the kernel, however they are in much better agreement. It is clear that the news noise model is much more constraining than the strain, which motivates using it for this work. This is in keeping with recent analyses in the ringdown literature \cite{2025arXiv250309678M, 2024arXiv241111269G}, although we note that the curvature scalar would also be an acceptable choice (for example, see \cite{2014PhRvD..90l4032L}).

%%%%%%%%%%%%%%%%%%%%
\end{document}